\documentclass[usenatbib]{solarphysics}
\usepackage[optionalrh,solaromanenum]{spr-sola-addons} % For Solar Physics 
\usepackage{epsfig}                     % For eps figures, old commands
\usepackage{graphicx}                    % For eps figures, newer & more powerfull
\usepackage{amssymb}                    % useful mathematical symbols
\usepackage{color}                       % For color text: \color command
\usepackage{url}                         % For breaking URLs easily through lines

\newcommand{\ssr}{{Space~Sci.~Rev.}}%
\newcommand{\grl}{{Geophys.~Res.~Lett.}}%
\newcommand{\jgr}{{J.~Geophys.~Res.}}%
\begin{document}

\begin{article}

\begin{opening}

\title{An Investigation of the CME of 3 November 2011 and its Associated Widespread Solar Energetic Particle Event}

%%%%%%%%%%%%%%%%%%%%%%%%%%%%%%%%%%%%%%%%%%%%%%%%%%%
%% Authors Names
%
\author{A.J.~\surname{Prise}$^{1}$\sep
        L.K.~\surname{Harra}$^{1}$\sep
        S.A.~\surname{Matthews}$^{1}$\sep
        D.M.~\surname{Long}$^{1}$\sep
        A.D.~\surname{Aylward}$^{2}$      
       }

%%%%%%%%%%%%%%%%%%%%%%%%%%%%%%%%%%%%%%%%%%%%%%%%%%%
%% Runningheads
%
\runningauthor{Prise et al.}
\runningtitle{A CME and its Associated Solar Energetic Particle Event}

%%%%%%%%%%%%%%%%%%%%%%%%%%%%%%%%%%%%%%%%%%%%%%%%%%%
%% Affilations 
%
  \institute{$^{1}$ UCL-Mullard Space Science Laboratory, Holmbury St. Mary, Dorking, Surrey, RH5 6NT, UK 
                     email: \url{ailsa.prise.11@ucl.ac.uk} \\ 
           $^{2}$ Department of Physics and Astronomy, UCL, Gower Street, London, WC1E 6BT, UK
                   \\
             }

%%%%%%%%%%%%%%%%%%%%%%%%%%%%%%%%%%%%%%%%%%%%%%%%%%%
%%% Abstract 
\begin{abstract}
Multi-spacecraft observations are used to study the \emph{in-situ} effects of a large CME erupting from the farside of the Sun on 3 November 2011, with particular emphasis on the associated solar energetic particle (SEP) event. At that time both \emph{Solar Terrestrial Relations Observatory} (STEREO) spacecraft were located more than 90 degrees from Earth and could observe the CME eruption directly, with the CME visible on-disk from STEREO-B and off the limb from STEREO-A. Signatures of pressure variations in the corona such as deflected streamers were seen, indicating the presence of a coronal shock associated with this CME eruption. The evolution of the CME and an associated EUV wave were studied using EUV and coronagraph images. It was found that the lateral expansion of the CME low in the corona closely tracked the propagation of the EUV wave, with measured velocities of 240$\pm$19 kms$^{-1}$ and 221$\pm$15 kms$^{-1}$ for the CME and wave respectively. Solar energetic particles were observed arriving first at STEREO-A, followed by electrons at the Wind spacecraft at L$_1$, then STEREO-B, and finally protons arriving simultaneously at Wind and STEREO-B. By carrying out velocity-dispersion analysis on the particles arriving at each location, it was found that energetic particles arriving at STEREO-A were released first and the release of particles arriving at STEREO-B was delayed by around 50 minutes. Analysis of the expansion of the CME to a wider longitude indicates that this delay is a result of the time taken for the edge of the CME to reach the footpoints of the magnetic-field lines connected to STEREO-B. The CME expansion is not seen to reach the magnetic footpoint of Wind at the time of solar particle release for the particles detected here, suggesting that these particles may not be associated with this CME.   
\end{abstract}

%%%%%%%%%%%%%%%%%%%%%%%%%%%%%%%%%%%%%%%%%%%%%%%%%%%
%% Keywords
%
%\keywords{}

\end{opening}
%-------------------------------------------------

%%%%%%%%%%%%%%%%%%%%%%%%%%%%%%%%%%%%%%%%%%%%%%%%%%%
%% Sections
%
\section{Introduction}\label{s:Intro}
Solar energetic particle events tend to fall into two broad categories: impulsive and gradual, as first described by \inlinecite{Reames1993}. Impulsive events have been identified from the acceleration of particles in a small region on the Sun, such as a solar flare, and produce short-duration events (several hours) over a narrow longitudinal range. Gradual events are commonly identified with acceleration from a wider source region, such as CME-driven coronal shocks or interplanetary shocks. These gradual events can have a duration of days and are often associated with Type II radio bursts. As noted by \inlinecite{Kahler1994} and \inlinecite{Tylka2003}, gradual event SEPs are released initially close to the Sun as the shock reaches around $\approx$3\,--\,10 solar radii. Once released, the accelerated particles propagate out along the magnetic-field lines, with the longitudinal spread of the resulting SEP event dependent on how broad the shock is. 

It was generally assumed that the angular spread of the SEPs was at least 100 degrees (\opencite{Cane2003}, \opencite{Kallenrode1993}) and \inlinecite{Cliver1995} used single-spacecraft observations to describe a farside flare with an observed coronal shock that resulted in an SEP event with a spread of at least 150 degrees in longitude implying a coronal shock that may extend to up to 300 degrees. 
Observations of coronal and interplanetary shocks have improved our understanding of wide SEP events. Interplanetary shocks can be as wide as 180 degrees in longitude (\opencite{Cliver1996},\opencite{Torsti1999}), which provides a very broad area for particle acceleration. Several signatures of shock formation in the corona have been identified, such as deflected streamers (\opencite{Cliver1999}; \opencite{Sheeley2000}) and the layer of electrons observed on the surface of the CME \cite{Vourlidas2003}. It has also been shown that these can be observed directly, along with pressure waves and shocks \cite{Manchester2008}. 

Previous investigators have shown that characteristics of SEP events such as particle intensities, anisotropies, and onset times can provide information about the magnetic-field line connections, \emph{e.g.} \inlinecite{Torsti2004}, \inlinecite{Bieber2005}. \inlinecite{Reames1996} showed that the time--intensity profile of gradual SEP events can provide information on the strength of the shock and therefore the strength of the connection with the source region. While particles accelerated by strong shocks are well connected to the source region and give a fairly constant profile, those accelerated by weaker shocks are poorly connected to the source region and yield a profile that decreases with time.

The launch of the \emph{Solar Terrestrial Relations Observatory} (STEREO: \opencite{Kaiser2008}) mission has provided a far better chance to study SEP events from multiple viewpoints. As the spacecraft separation increases, it has become evident that SEP events can be detected over a far wider range of longitudes than previously assumed, such as the event described by \inlinecite{Dresing2012}, where a backsided source region produced an SEP event with a longitudinal spread of up to 300 degrees. \inlinecite{Rouillard2012} describe a wide SEP event detected at L$_1$ and at STEREO-A (at this time separated by 88$^{\circ}$), caused by a fast and wide CME. They observe a delay in the particle release time of particles arriving at L$_1$, which they conclude is due to the time taken for the CME to expand to the longitude of the magnetic footpoint connected to L$_1$.

\begin{figure}[t]
\centering
\caption{When different \emph{in-situ} signatures that are detected at each spacecraft. The start time of the plot is 22:10 UT, the time of the CME eruption.}
\includegraphics[width=0.69\textwidth,clip=]{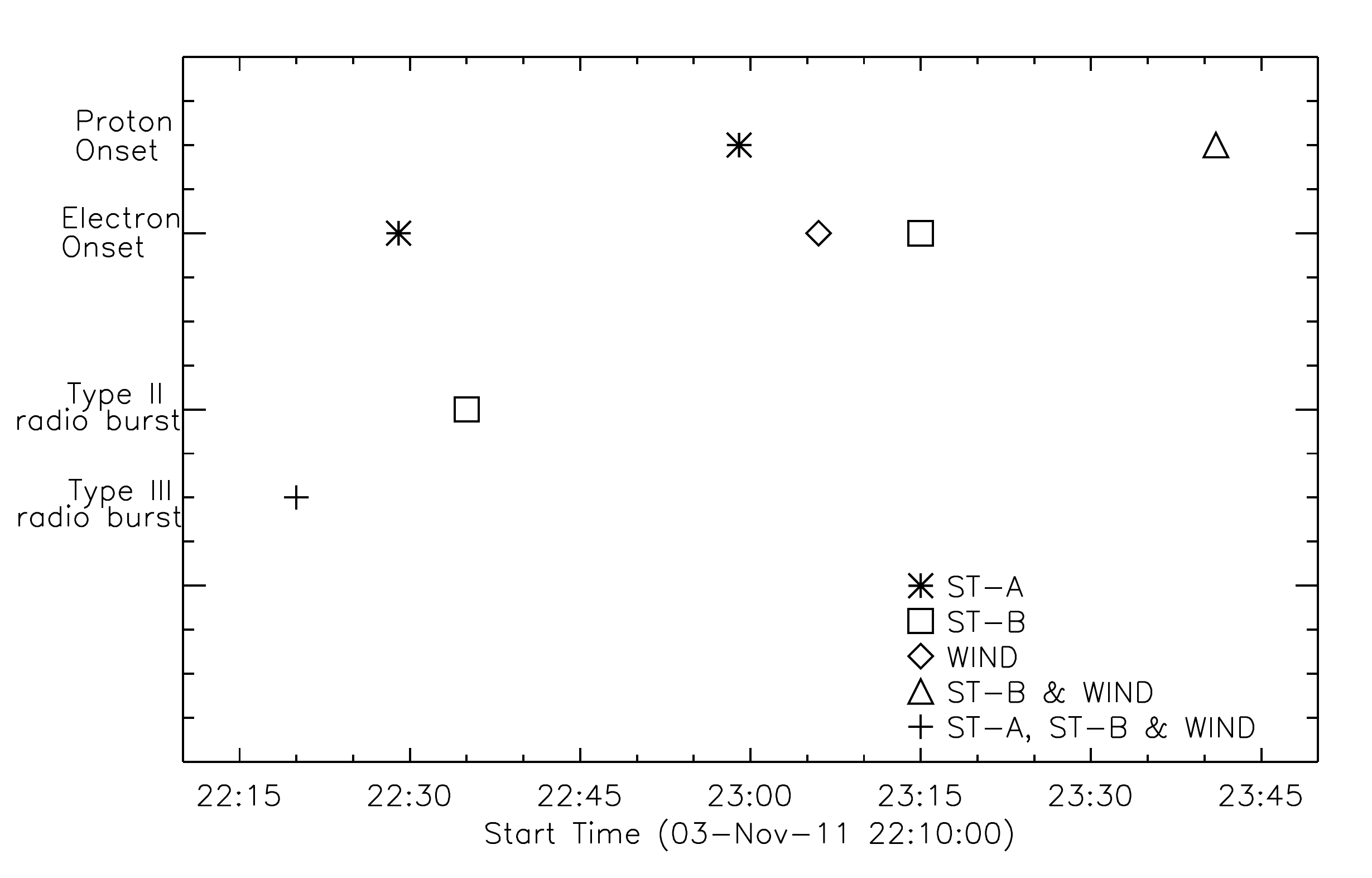}
\label{fig:timeline}
\end{figure}

\begin{figure}[t,b]
\centering
\caption{Positions of STEREO-A, STEREO-B, and the Earth on 3 November 2011 22:00 UT. The direction of the erupting CME is indicated by the black arrow and the Parker spiral is also overlaid. This is a modified version of the plot found from \protect\url{http://stereo-ssc.nascom.nasa.gov/cgi-bin/make_where_gif}.}
\includegraphics[width=0.6\textwidth,clip=]{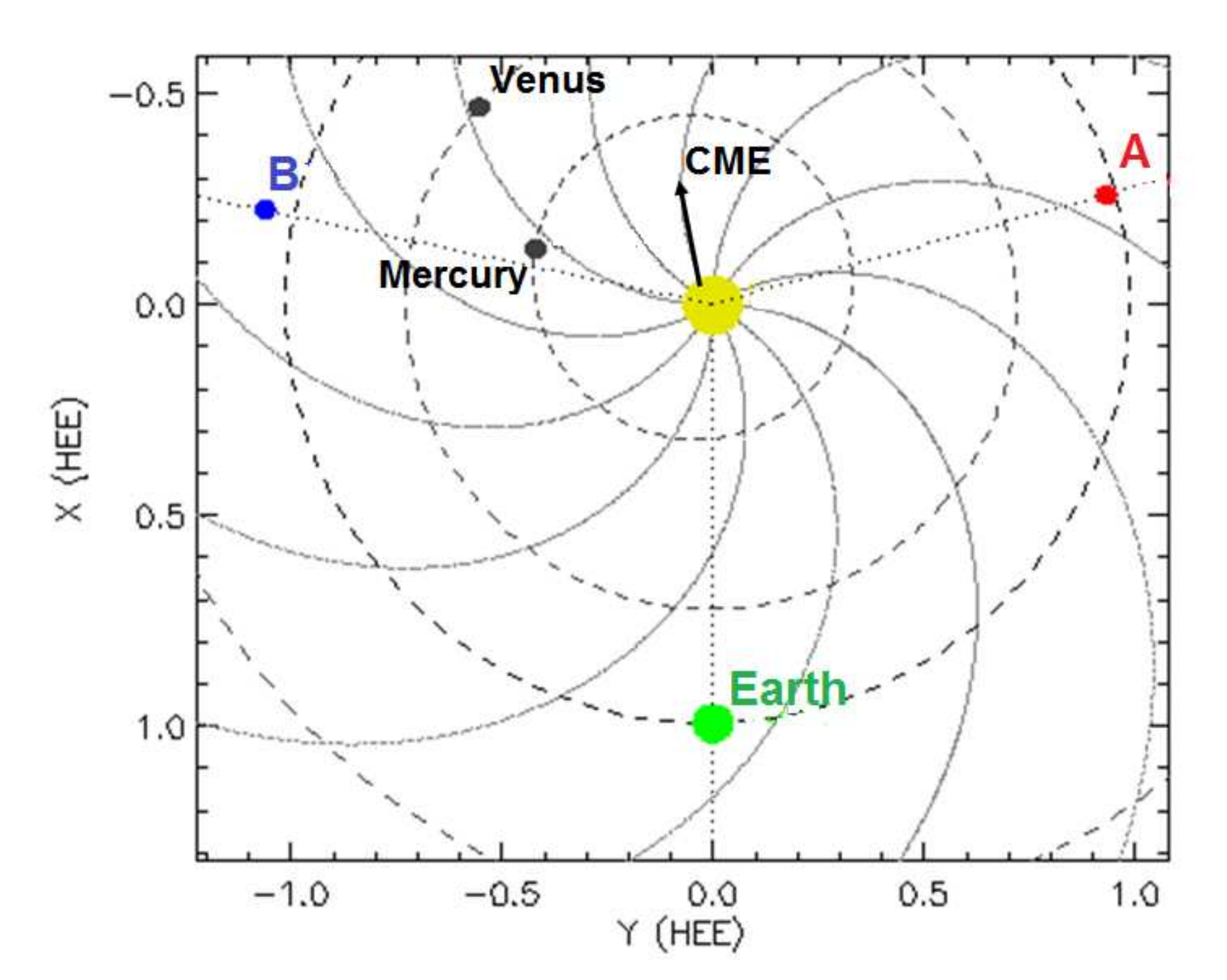}
\label{fig:CMEdirn}
\end{figure}

In the present study, the characteristics of a CME that erupted from the farside of the Sun on 3 November 2011 are analysed. This CME was accompanied by an EUV wave, and produced several \emph{in-situ} signatures  at 1 AU, including energetic particles and radio bursts. A summary of the timings of the CME and \emph{in-situ} signatures is shown in Table \ref{tbl:timings} and Figure \ref{fig:timeline}. At this time the STEREO spacecraft were past quadrature and were separated by 152$^{\circ}$, 106$^{\circ}$ and 102$^{\circ}$ ahead and behind the Earth respectively. Figure \ref{fig:CMEdirn} shows the positions of STEREO-A, STEREO-B, and the Earth, along with the direction of the erupting CME.
The evolution of the CME and its associated EUV wave are analysed here, allowing the arrival times of the energetic particles at STEREO-A and STEREO-B to be explained.

%% Table
%
 \begin{table}[t]
 \caption{The detected onset times of various \emph{in-situ} signatures studied, where they are detected, and where possible, the release time from the Sun.}\label{tbl:timings}
 \begin{tabular}{cccc}     
 \hline
Time [UT] & Event & Position & Release Time [UT] \\ \hline
 22:10 & CME and EUV wave & backsided, visible from ST-B \\
22:20 & Type III radio burst & ST-A, -B $\&$ Wind & 22:12 \\ 
22:29 & Electron onset & ST-A \\ 
22:35 & Type II radio burst & ST-B & 22:27 \\ 
22:59 & Proton onset & ST-A & 22:19$\pm$00:14 \\
23:06 & Electron onset & Wind & 22:47$\pm$00:15 \\ 	
23:15 & Electron onset & ST-B \\ 
23:41 & Proton onset & ST-B & 23:00$\pm$00:08 \\ 
23:41 & Proton onset & Wind \\ \hline
 \end{tabular}
 \end{table}

\section{Instrumentation}
For this study, data were utilised from multiple spacecraft and instruments in order to investigate the global influence of the CME and its impacts \emph{in-situ}. The CME eruption was observed directly by the \emph{Sun--Earth Connection Coronal and Heliospheric Investigation} (SECCHI: \opencite{Howard2008}) instrument suite on STEREO; the \emph{Extreme Ultraviolet Imager} (EUVI: \opencite{Wuelser2004}) was used to study the solar disk, with the CME propagation observed using the  Cor-1 and Cor-2 (\opencite{Thompson2003}) coronagraphs. The erupting CME was also tracked using the \emph{Large Angle and Spectrometric Coronagraph} experiment (LASCO: \opencite{Brueckner1995}) onboard the \emph{Solar and Heliospheric Observatory} (SOHO: \opencite{Domingo1995}). 

\emph{In-situ} measurements were taken using the \emph{In-situ Measurements of Particles and CME Transients} (IMPACT: \opencite{Luhmann2008}) package of instruments on the STEREO spacecraft. This includes the \emph{High Energy Telescope} (HET: \opencite{vonRosenvinge2007}), \emph{Low Energy Telescope} (LET: \opencite{Mewaldt2008}), and \emph{Solar Electron and Proton Telescope} (SEPT: \opencite{Muller-Mellin2008}) for energetic particle measurements. The SWAVES instrument (\opencite{Bougeret2008}) on--board the STEREO spacecraft was used to measure radio bursts, along with the equivalent instrument WAVES (\opencite{Bougeret1995}) on the Wind spacecraft at L$_1$. Wind also provided energetic particle measurements at L$_1$ with the \emph{Energetic Particles: Acceleration, Composition and Transport instrument} (EPACT: \opencite{vonRosenvinge1995}) for proton measurements and with the \emph{3D Plasma Analyzer} (3DP: \opencite{Lin1995}) for electron measurements.

\section{Observations}
\begin{figure}[t]
\centering
\caption{Running-difference images from Cor2-B, C2 (LASCO), and Cor2-A showing the eruption of the CME as well as deflected streamers. The streamers appear in these running-difference images, indicating that they have moved from their previous position and have therefore been deflected.}
\setlength{\tabcolsep}{0.1pt}
\begin{tabular}{ccc}
\includegraphics[width=0.323\textwidth,clip=]{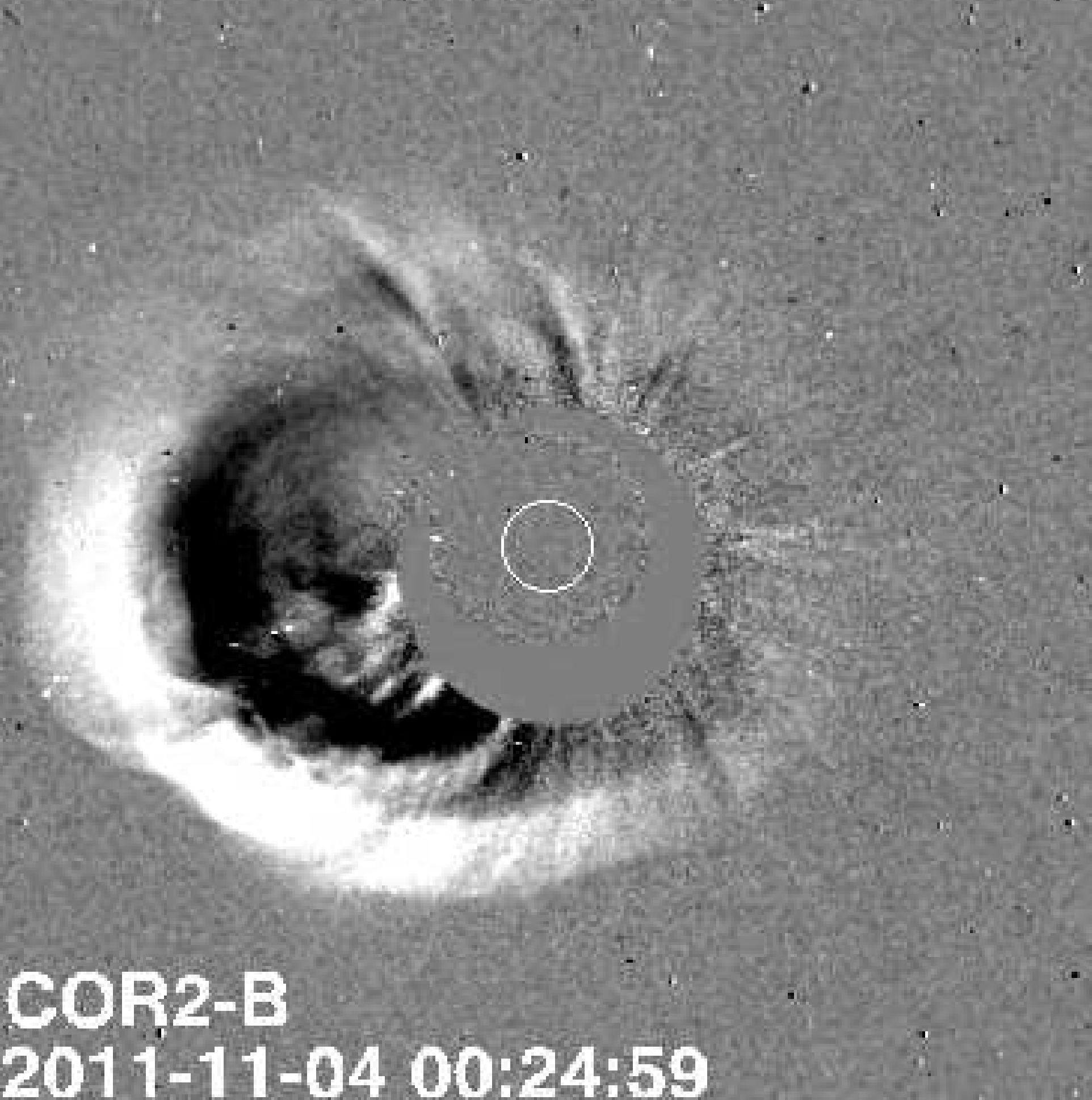}&
\includegraphics[width=0.323\textwidth,clip=,angle=90]{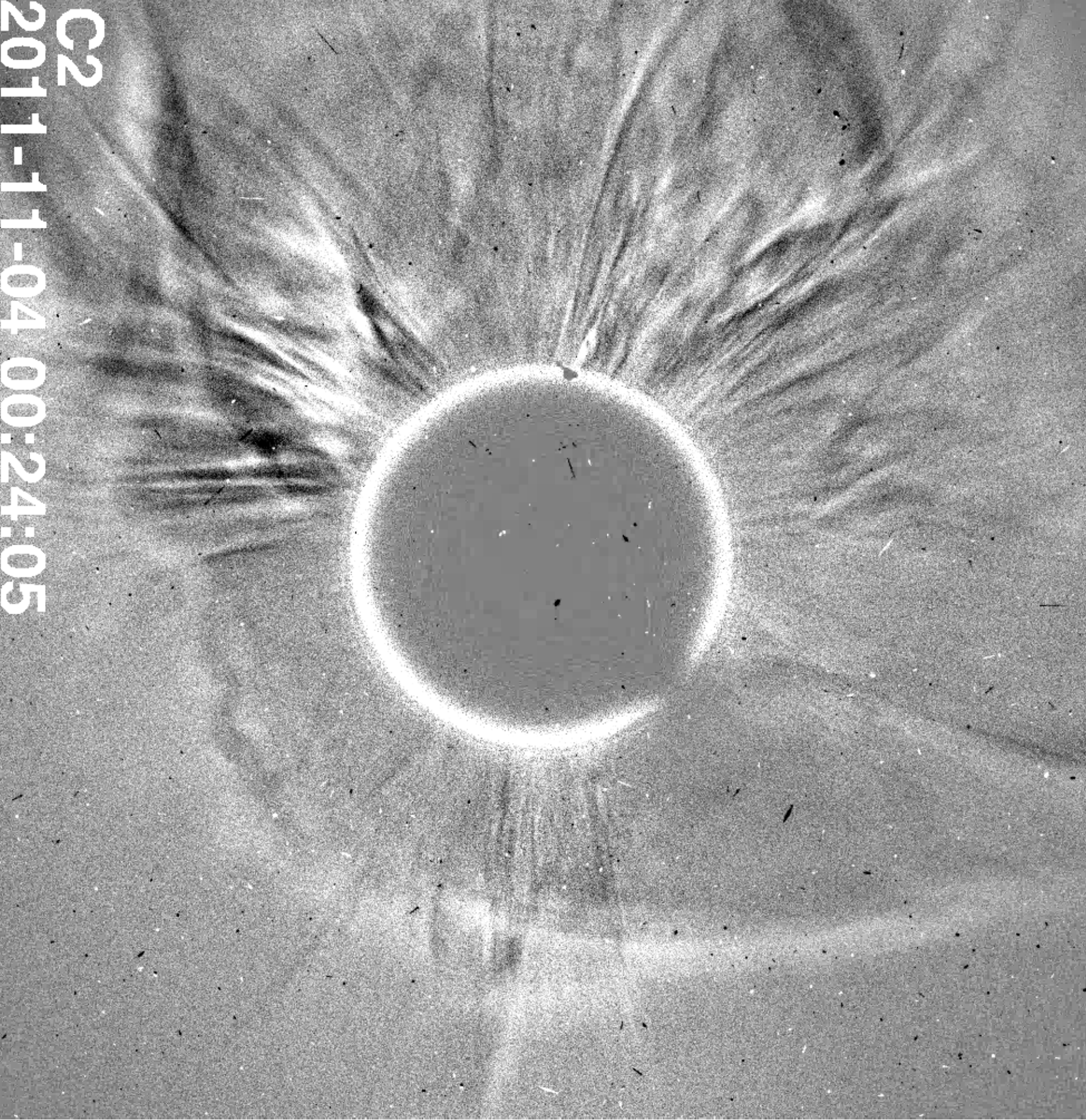}&
\includegraphics[width=0.323\textwidth,clip=,angle=90]{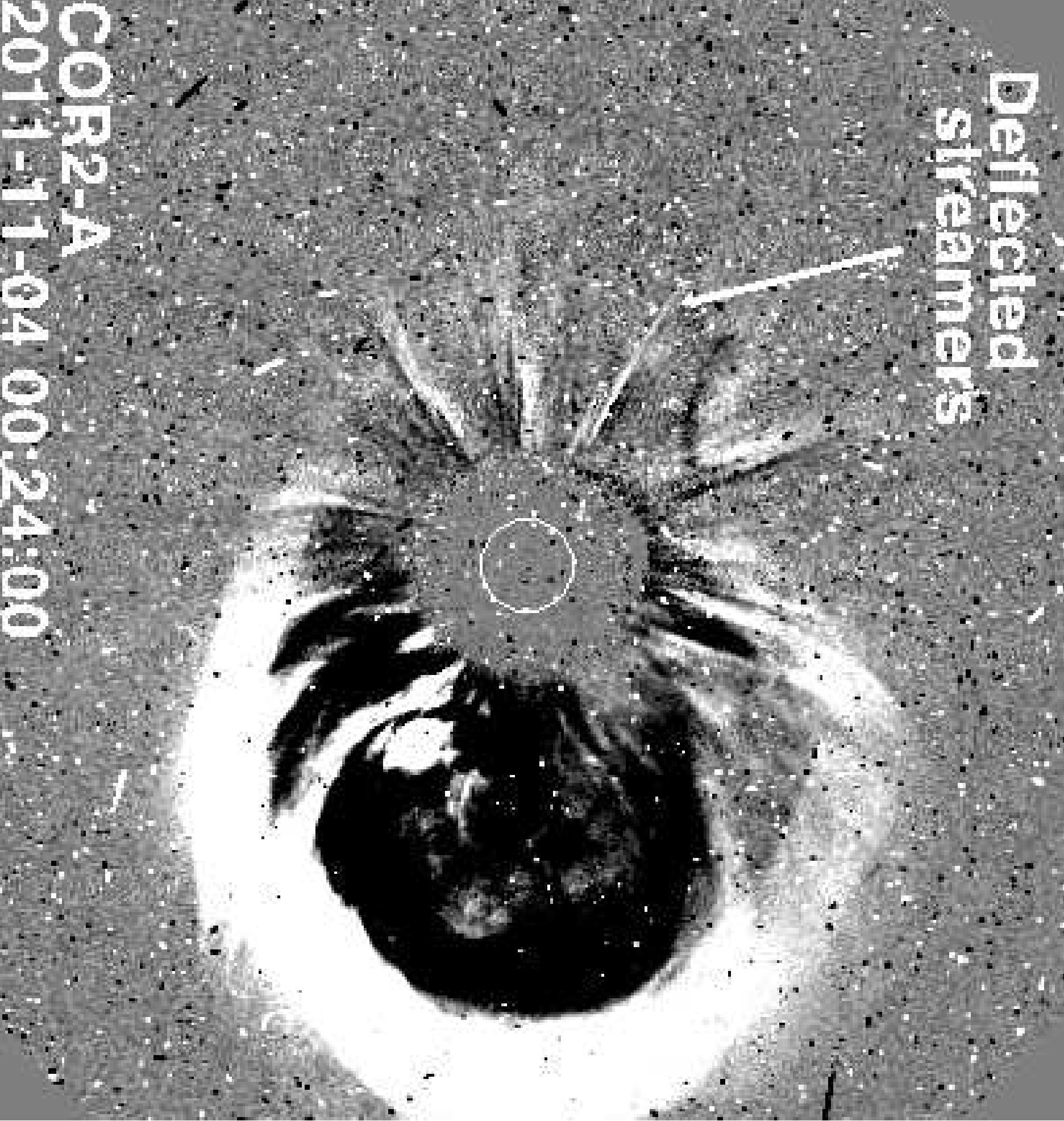}
\end{tabular}
\label{fig:cor}
\end{figure}

The CME was observed on the disk by STEREO-B and on the western limb of STEREO-A, with the eruption beginning at about 22:10 UT on 3 November 2011. In coronagraph  images, it is seen erupting off the western limb of STEREO-A and off the eastern limb of STEREO-B, indicating that the CME originated at the farside of the Sun from the Earth. It is also observed from LASCO, erupting off the eastern limb, consistent with observations of the erupting active region on-disk in STEREO-B (see Figure \ref{fig:cor}). Height--time plots of STEREO-A coronagraph observations give this CME a radial speed of 972 kms$^{-1}$, which should suffer from very small projection effects as the CME erupts off the limb as viewed from STEREO-A. 

\begin{figure}[t]
\caption{Radio measurements made by SWAVES at STEREO-A (top), STEREO-B (middle) and WAVES on WIND at L$_1$ (bottom). The Type II radio burst is visible in the middle panel (STEREO-B) from approximately 22:35\,--\,00:10 UT. Type III radio bursts were also observed at all three locations at around 22:20 UT.}\label{fig:radioplot}
\centerline{\includegraphics[width=0.85\textwidth,clip=]{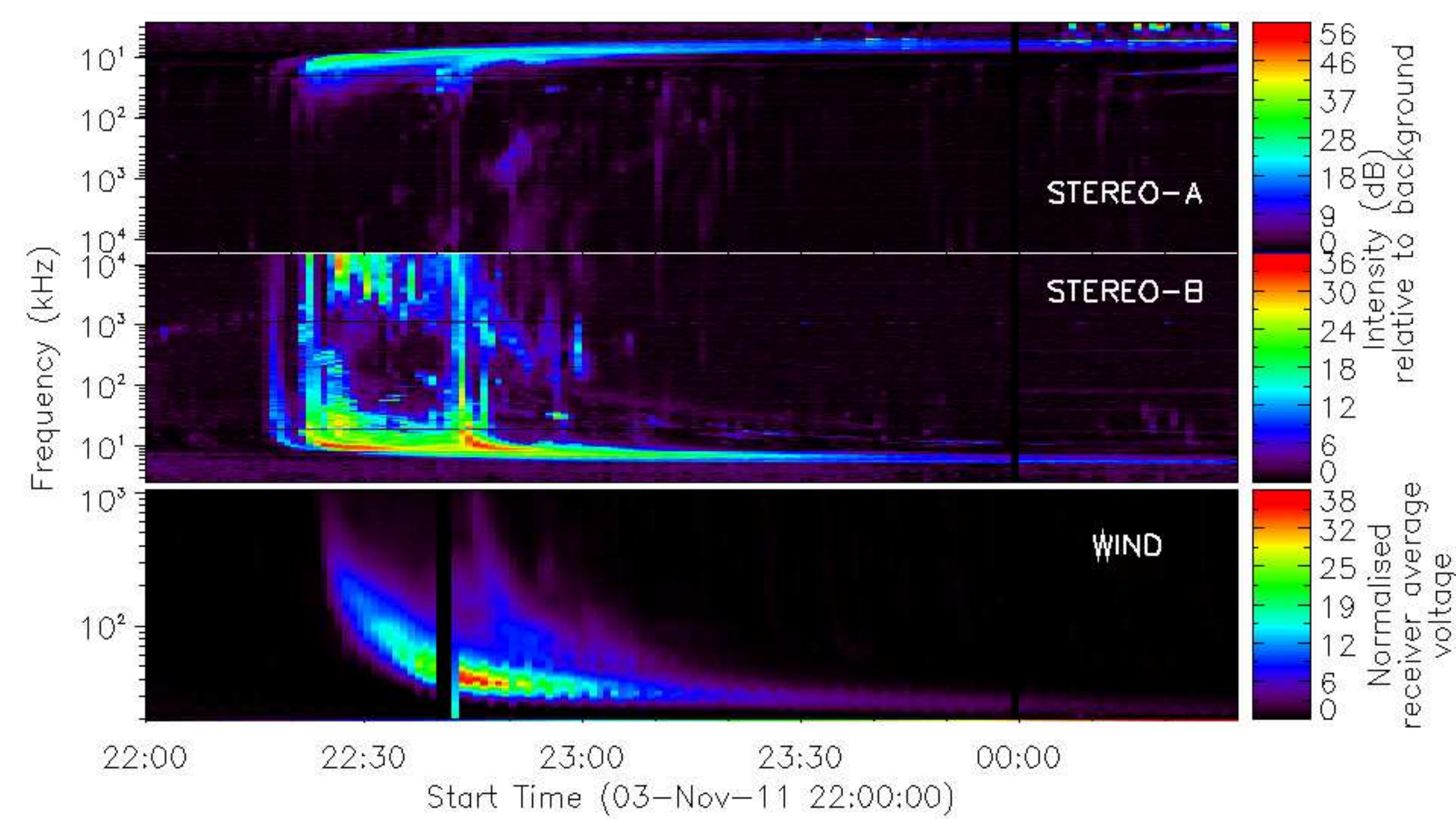}}
\end{figure} 

Coronagraph observations of this CME show several signatures of pressure variations, including deflected streamers. These are observed as streamers appearing in running-difference coronagraph images, as this shows that the streamers have moved from their previous position. These deflected streamers appear in the coronagraph images at roughly 00:24 UT, on the opposite limb to the CME as seen in Figure \ref{fig:cor}. This suggests that the CME has an influence that extends to the other side of the Sun. Figure \ref{fig:radioplot} shows the radio measurements around the time of this CME. A Type II radio burst, indicative of a shock, is detected by SWAVES-B between 22:35 UT and around 00:00 UT. A Type III radio burst is also visible at all three locations, at around 22:20 UT.

\subsection{Propagation of the EUV Wave and Lateral CME Expansion}
The EUV wave associated with this eruption was observed from STEREO-B between 22:16 and about 22:31 UT, travelling from the erupting active region, near the eastern limb of STEREO-B, towards the middle of the disk. The direction of the propagation of the wave and position of the erupting active region meant that it was not observed by STEREO-A. 

\begin{figure}[t]
\centering
\caption{Percentage base-difference images of the EUV wave in 195\,\AA\ seen from 22:16 to 22:31 UT from STEREO-B. The white dashed lines indicate the region which was used to generate the intensity profiles.}
\includegraphics[width=0.85\textwidth,clip=]{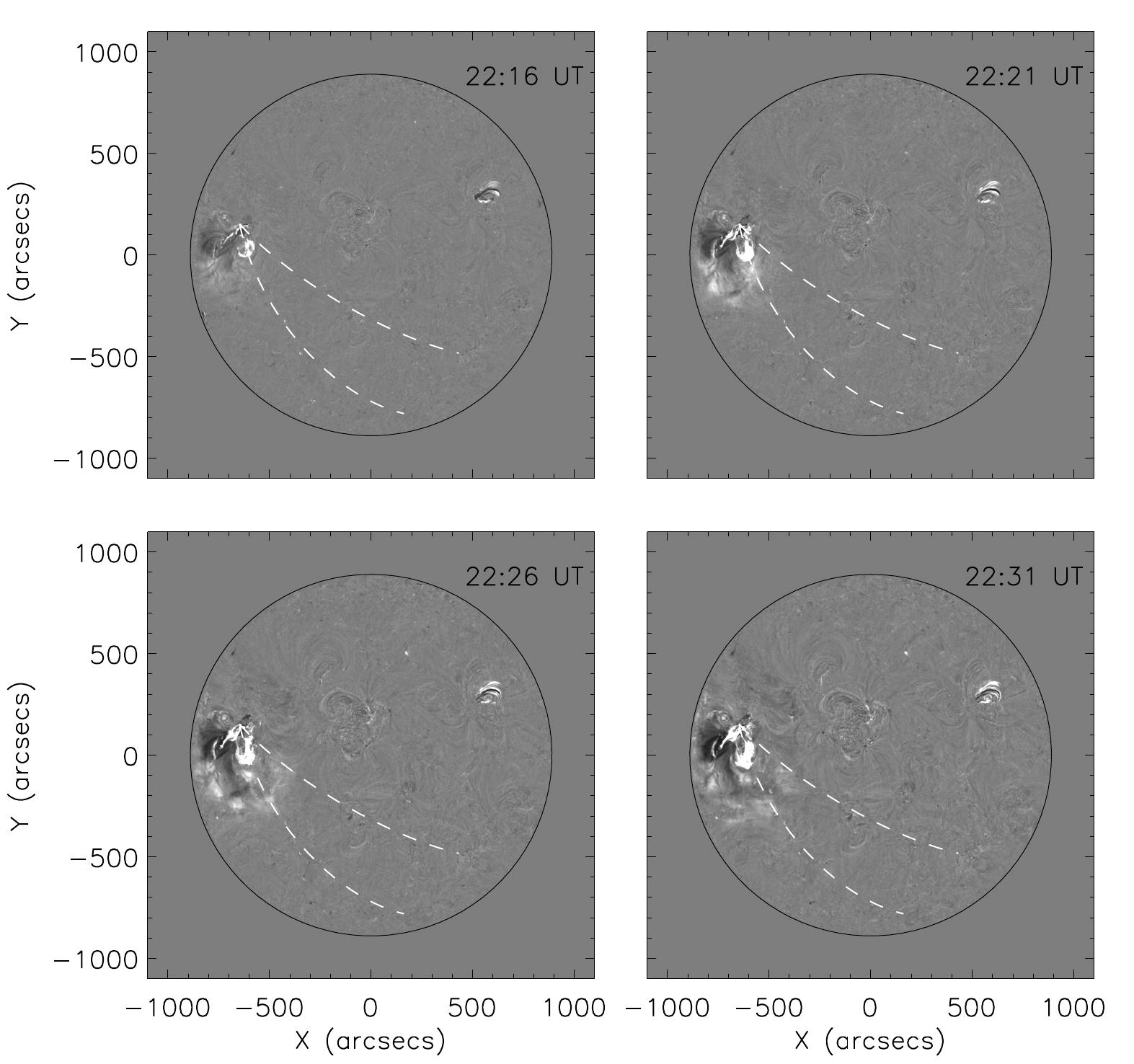}\\
\label{fig:pbd_map}
\end{figure}

\begin{figure}[t]
\centering
\caption{Intensity profiles for the EUV wave (crosses) with the Gaussian fit of the wave over-plotted.}
\includegraphics[width=0.7\textwidth,clip=]{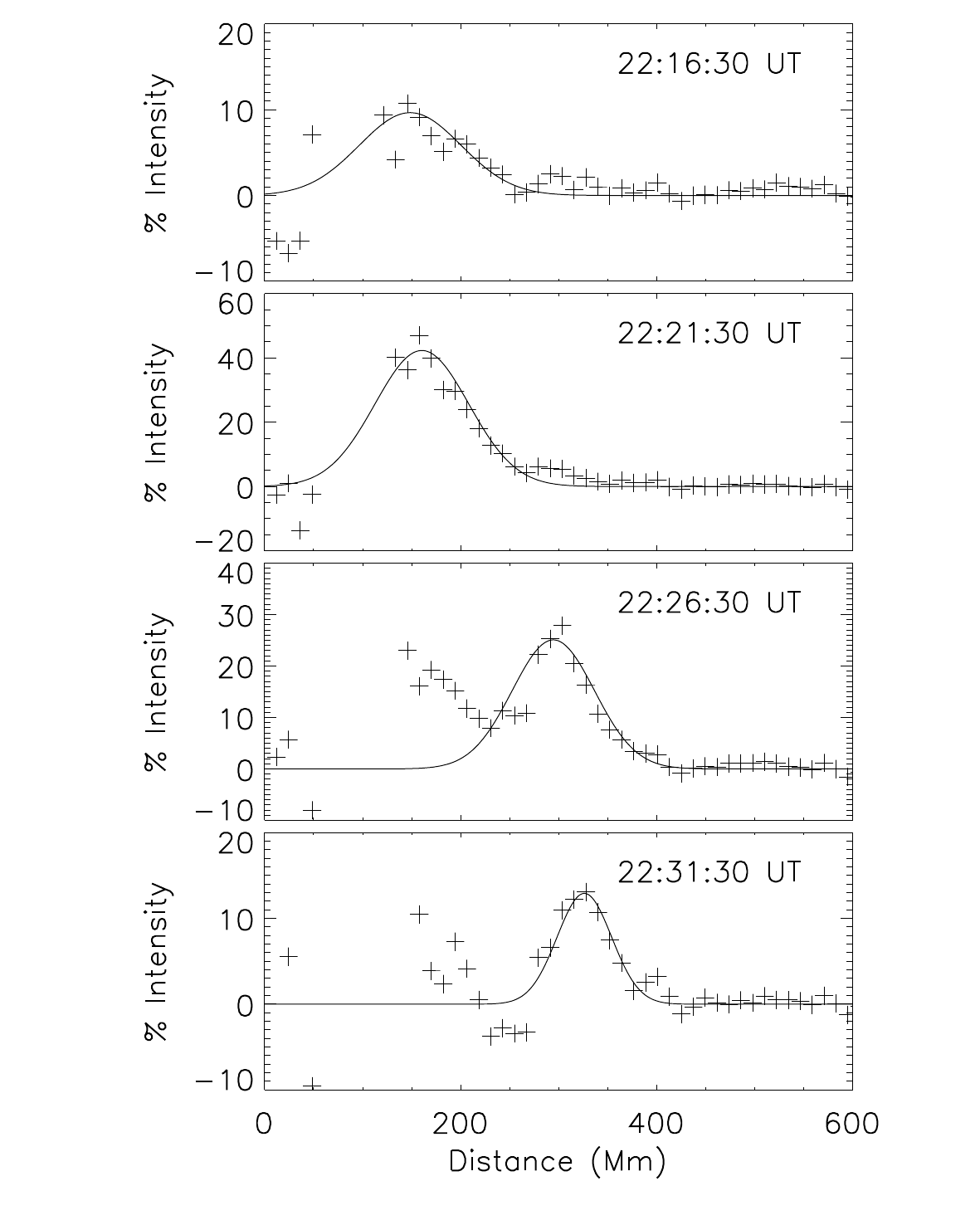}
\label{fig:intensityplot}
\end{figure}

\begin{figure} [h,t,b]
\centerline{\includegraphics[width=0.8\textwidth,clip=]{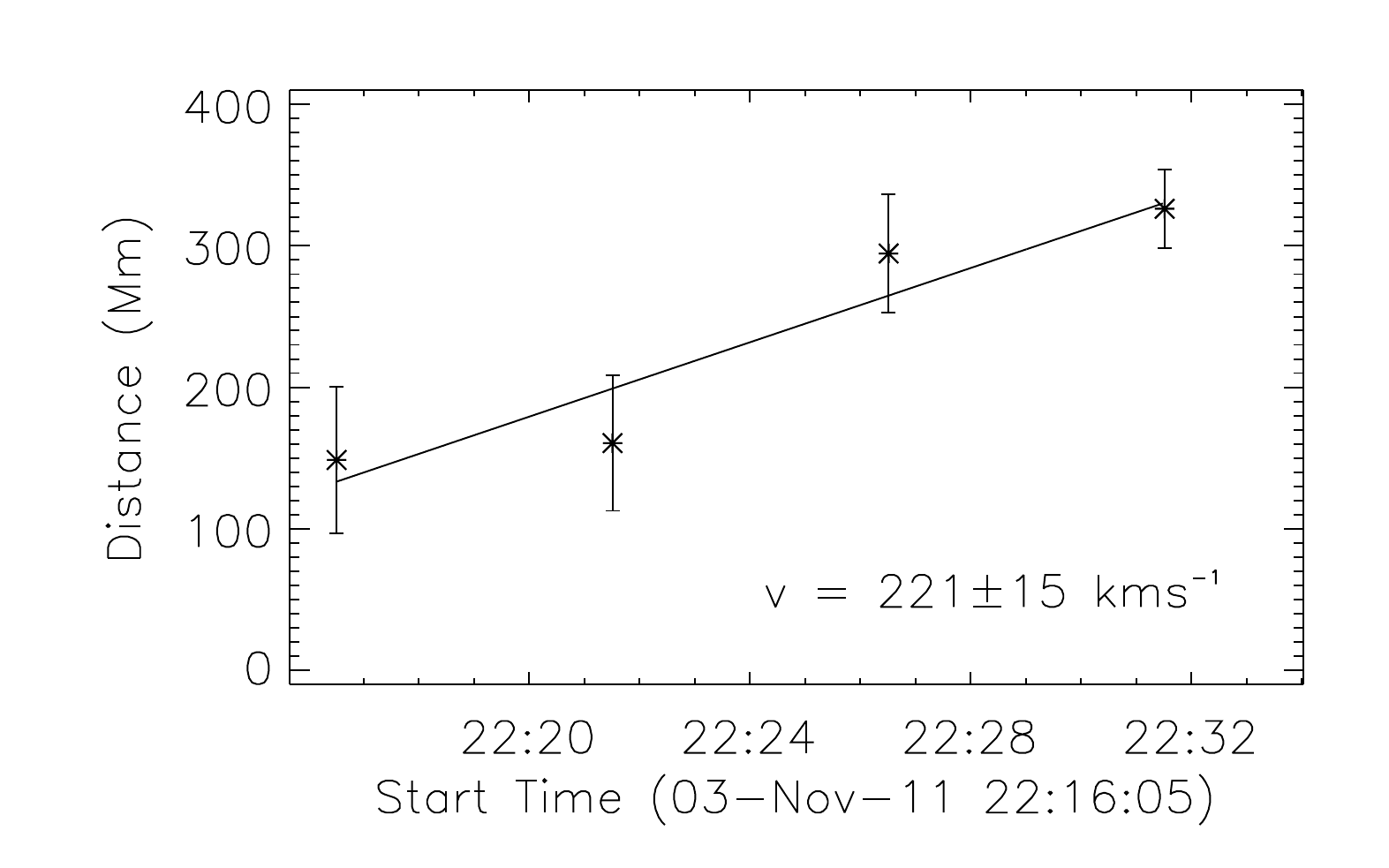}}
 \caption{Distance--time plot of the EUV wave. The distance is measured from the source point of the wave, defined as the flare position. v is the velocity of the wave found from this plot. The error bars are the FWHM of the Gaussian fit of the wave pulse from the intensity plots.}\label{fig:wavedistanceplot}
\end{figure}

The propagation of the wave was measured using an intensity-profile technique (\emph{cf.} \opencite{Long2011}, \opencite{Muhr2011}, \opencite{Long2011a}). This method calculates the intensity of percentage base-difference images of the Sun in a defined arc sector from the source of the wave, as shown in Figure \ref{fig:pbd_map}. 
The intensity is measured in several possible arc sectors and then the specific arc sector chosen that detects the maximum number of pulse identifications, in order to better track the propagation of the wave and ensure sufficient detections to allow determination of accurate kinematics. A Gaussian model is then fitted to the resulting intensity profile, allowing the determination of the position of the pulse with time (Figure \ref{fig:intensityplot}). The error in these positions is the full width at half maximum of the Gaussian fitted to that pulse. The distance can then be plotted against the time to find the wave speed, as shown in Figure \ref{fig:wavedistanceplot}. This method only picked up the wave between 22:16 and 22:31 UT, propagating a distance of about 200 Mm in that time. The wave speed was found to be 221$\pm$15 kms$^{-1}$, where the error is due to the uncertainty in the fit of the distance--time plot.

\begin{figure}[t,b]
\centering
 \caption{A combined EUVI and Cor-1 polar plot. The lateral expansion speed of the CME is calculated from the changing position angle of the edge of CME with time.}
\includegraphics[width=0.9\textwidth,clip=]{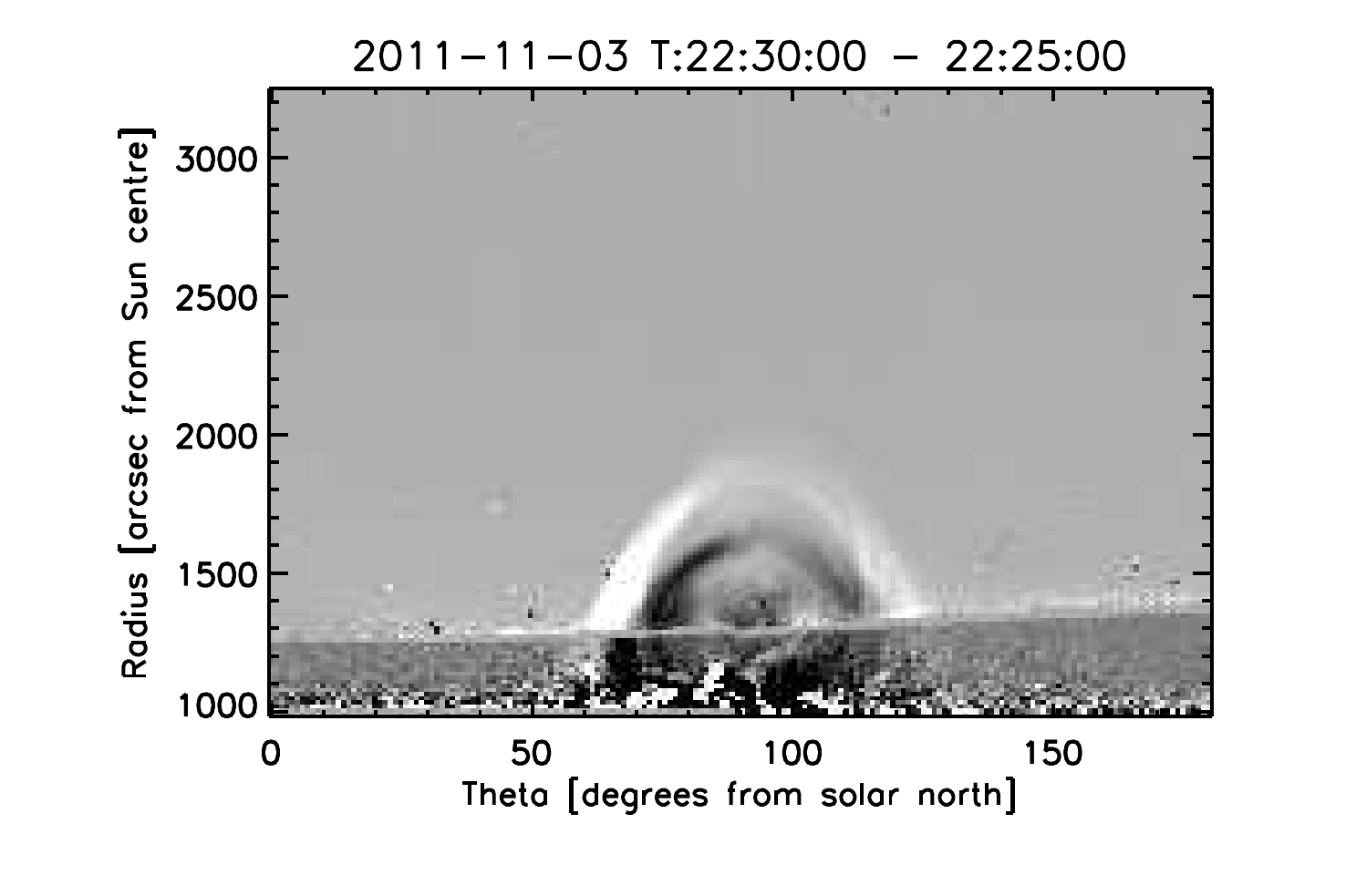}
\label{fig:polarplot}
\end{figure}

\begin{figure}
\centering
\caption{Distance-time plot of the lateral expansion of the CME as measured from polar plots of EUVI (left) and Cor-1 (right). The distance is determined using polar plots such as those in Figure \ref{fig:polarplot} and the velocities found are 240$\pm$19 kms$^{-1}$ for EUVI and 674$\pm$38 kms$^{-1}$ for Cor-1.}
\setlength{\tabcolsep}{0.5pt}
\begin{tabular}{cc}
\includegraphics[width=0.49\textwidth,clip=]{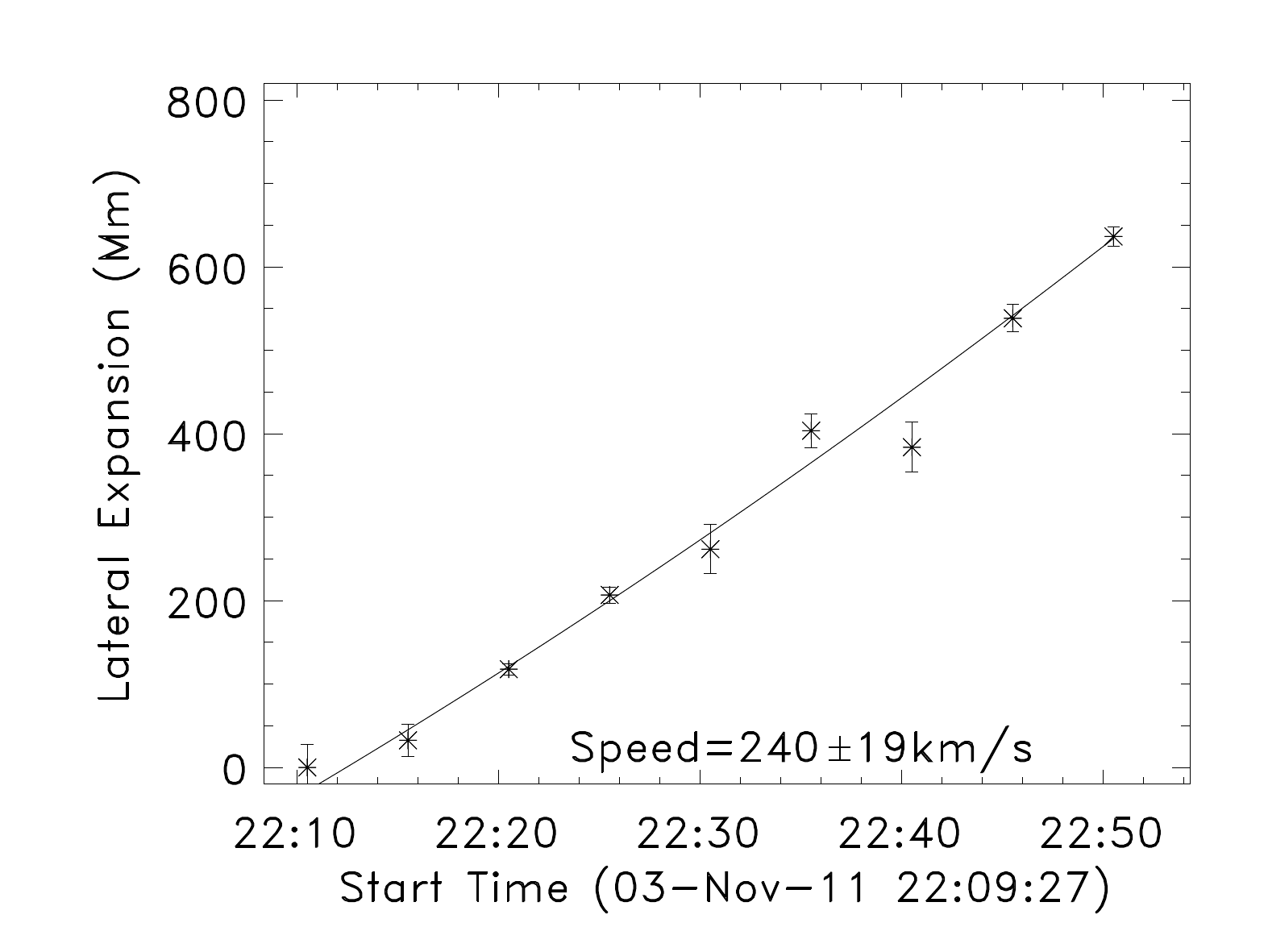}&
\includegraphics[width=0.49\textwidth,clip=]{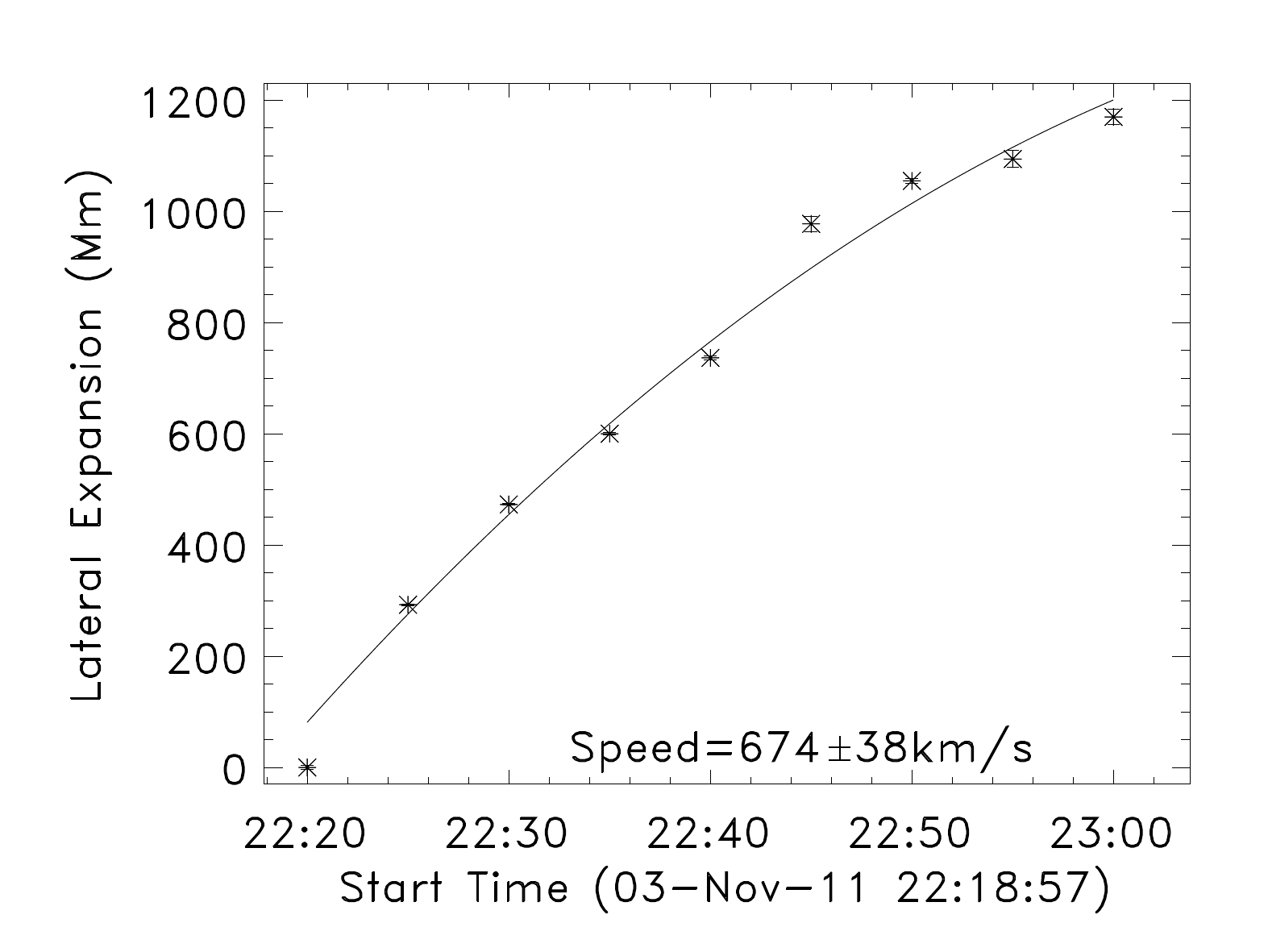}\\
\end{tabular}
\label{fig:latexppolarnew}
\end{figure}

\begin{figure} [h,t,b]
\centering
\caption{Combined EUVI-B and Cor1-B running-difference images showing that the EUV wave visible in EUVI seems to track the expansion of the CME}
\setlength{\tabcolsep}{0.5pt}
\begin{tabular}{ccc}
\includegraphics[width=0.325\textwidth,clip=]{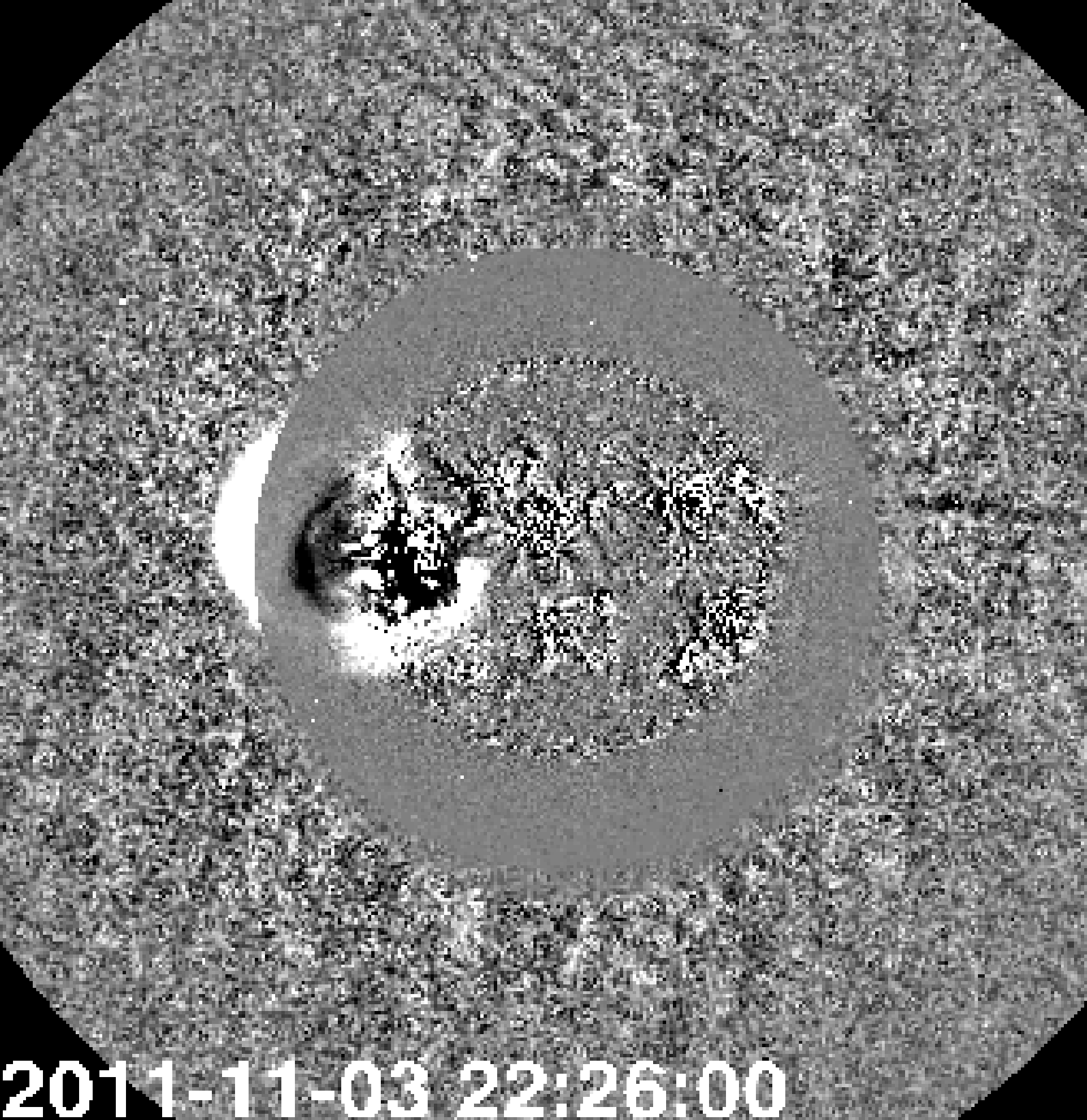}&
\includegraphics[width=0.325\textwidth,clip=]{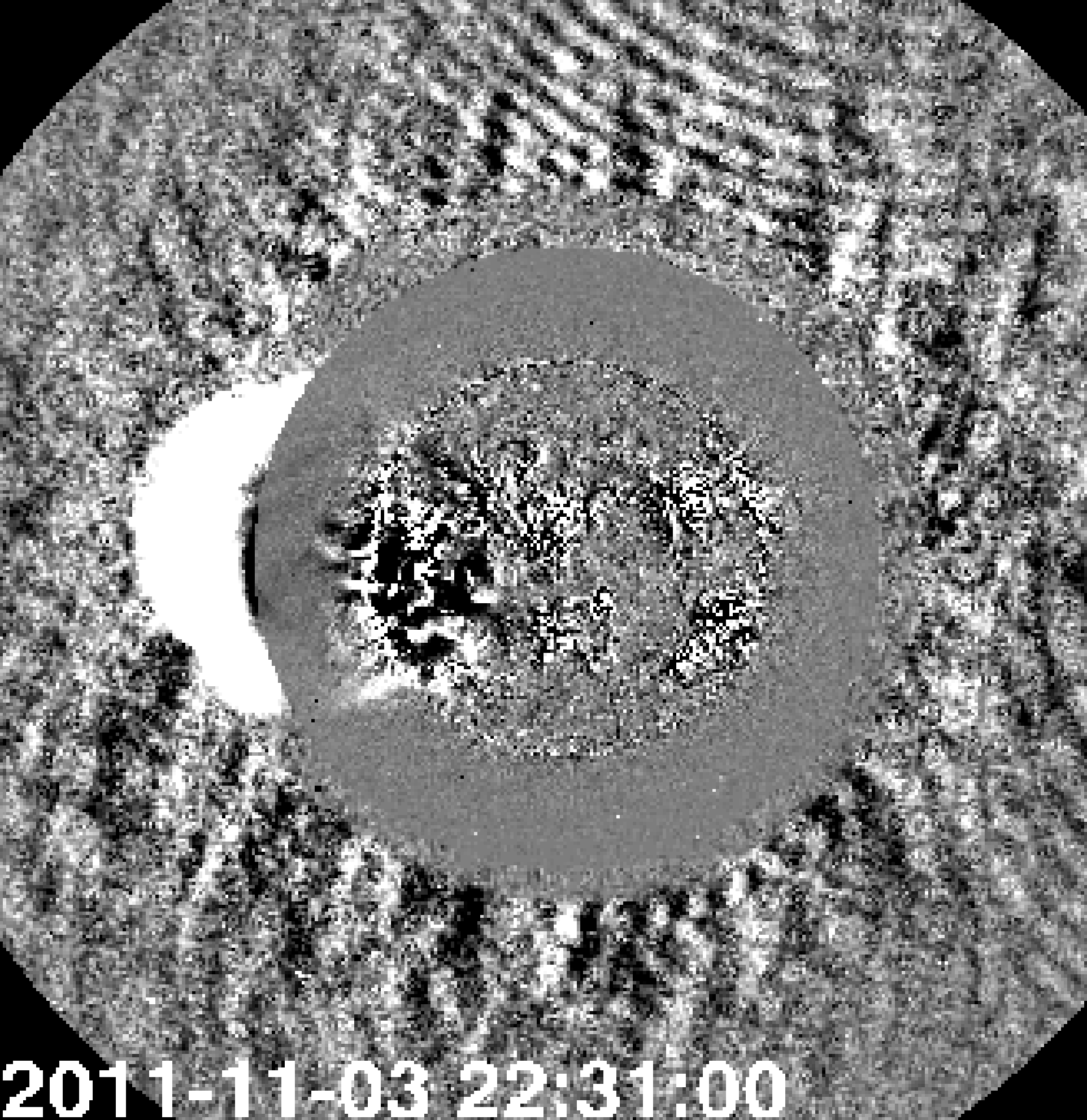}&
\includegraphics[width=0.325\textwidth,clip=]{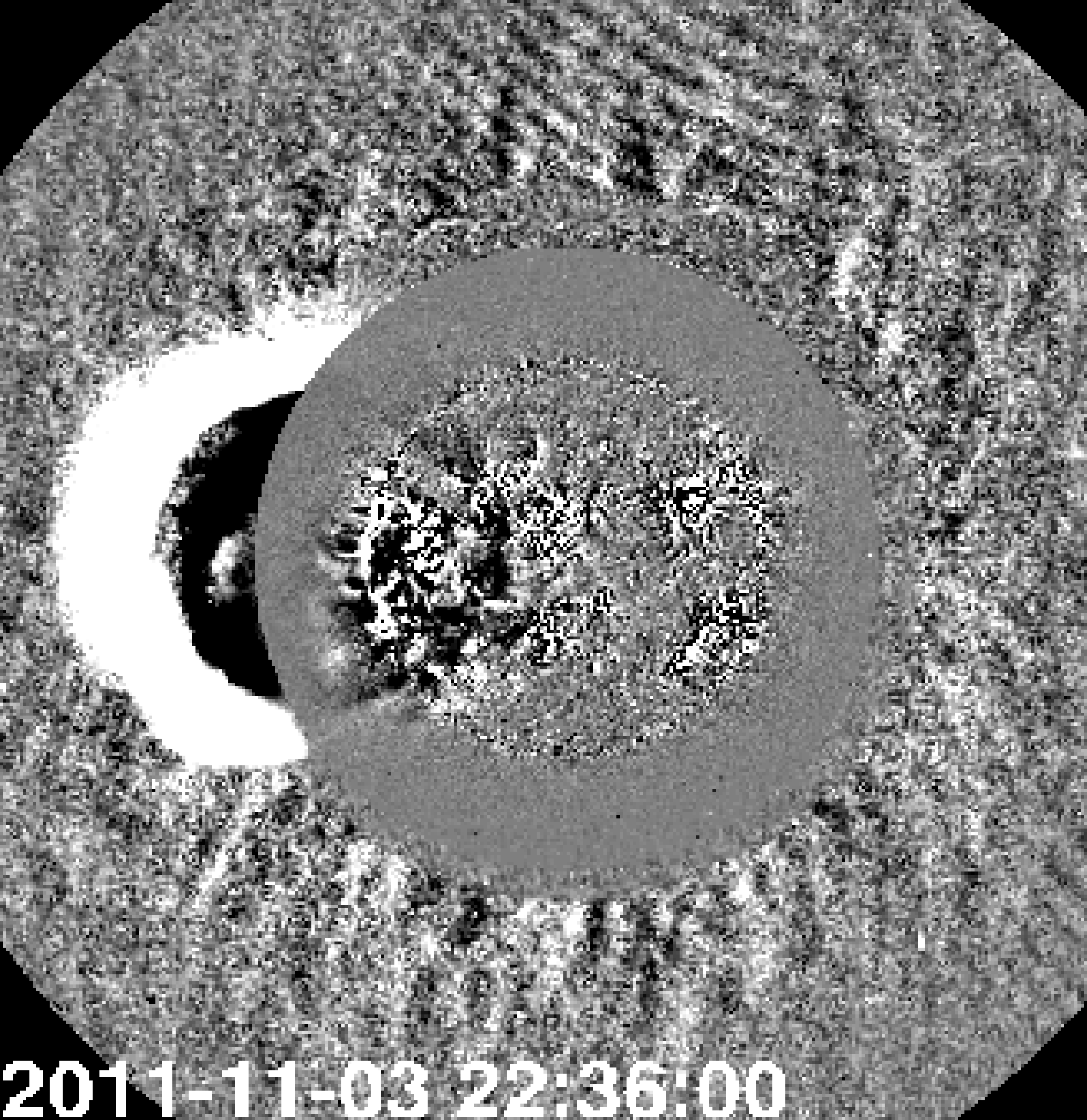}
\label{fig:euvicor1brdiffcomb}
\end{tabular}
\end{figure}

The lateral expansion of the CME was determined by tracking the CME edge using polar plots of EUVI and Cor-1 images (see Figure \ref{fig:polarplot}). The expansion of the CME lower down in the corona was tracked between 22:10 and 22:50 UT by measuring the changing position angle of the edge of the CME over time, as seen in the polar plots. From this, the distance that the CME has expanded is calculated and the velocity found from a distance--time plot, shown in Figure \ref{fig:latexppolarnew}. In this time the CME expanded by roughly 600\,Mm. The position angle was measured three times and the standard error in these three independent measurements was used as to determine the uncertainty in the distance expanded by the CME. The velocity of this expansion at the level of EUVI was found to be 240$\pm$19 kms$^{-1}$, where the error comes from the uncertainty in the fit of the distance--time plot. This velocity suggests that the expansion of the CME initially roughly tracks the propagation of the EUV wave, as this speed is consistent with that of the wave within errors and the CME edge appears to track the EUV wave in images, as shown in Figure \ref{fig:euvicor1brdiffcomb}. The lateral expansion was also measured in Cor-1 between 22:20 and 23:00 UT, using the same method with Cor-1 polar plots. The CME was seen to expand by roughly 1200\,Mm at this height and the velocity was found to be significantly higher than the expansion speed from EUVI polar plots, at 674$\pm$38 kms$^{-1}$.

\subsection{Solar Energetic Particle Events}
Solar energetic particles were detected at both STEREO spacecraft and at Wind after the eruption of this CME. Energetic electrons arrived first at STEREO-A at 22:29$\pm$00:09 UT followed by protons 30 minutes later. Accelerated electrons were detected at Wind at 23:06$\pm$00:09 UT and STEREO-B at 23:15$\pm$00:07 UT, and energetic protons then arrived at Wind and STEREO-B simultaneously at 23:41 UT.

The onset times were determined using the method described by \inlinecite{Krucker1999}, where the background flux is subtracted and the curves normalised in units of standard deviation. An upper limit for the onset time is then defined as the time when this normalised flux first goes above 4$\sigma$. The actual onset time is then taken to be the time increment after the latest time that the normalised flux is negative prior to the upper limit. The difference between this onset time and the upper limit is then the uncertainty in the onset time.

Energetic proton measurements were taken at the STEREO spacecraft using the HET and LET instruments in three energy bins each between 13 and 100  MeV for HET (13\,--\,21 MeV, 21\,--\,40 MeV and 40\,--\,100 MeV), and 2.2 and 12 MeV for LET (2.2\,--\,4 MeV, 4\,--\,6 MeV and 6\,--\,12 MeV). The proton measurements at L$_1$ were taken with the EPACT instrument on the Wind spacecraft in two energy bins between 19 and 72 MeV (19\,--\,28 MeV and 28\,--\,72 MeV). Energetic electron measurements at the STEREO spacecraft use the SEPT instrument, measuring electrons in four energy bins between 0.03 and 0.47 MeV (0.03\,--\,0.06 MeV, 0.06\,--\,0.11 MeV, 0.11\,--\,0.24 MeV and 0.24\,--\,0.47 MeV). The electron measurements at L$_1$ use the 3DP instrument on the Wind spacecraft, which measures energetic electrons in three energy bins between 0.02 and 0.225 MeV (0.02\,--\,0.048 MeV, 0.043\,--\,0.138 MeV and 0.127\,--\,0.225 MeV).

 \begin{figure}[t,b]
\centerline{\includegraphics[width=0.9\textwidth,clip=]{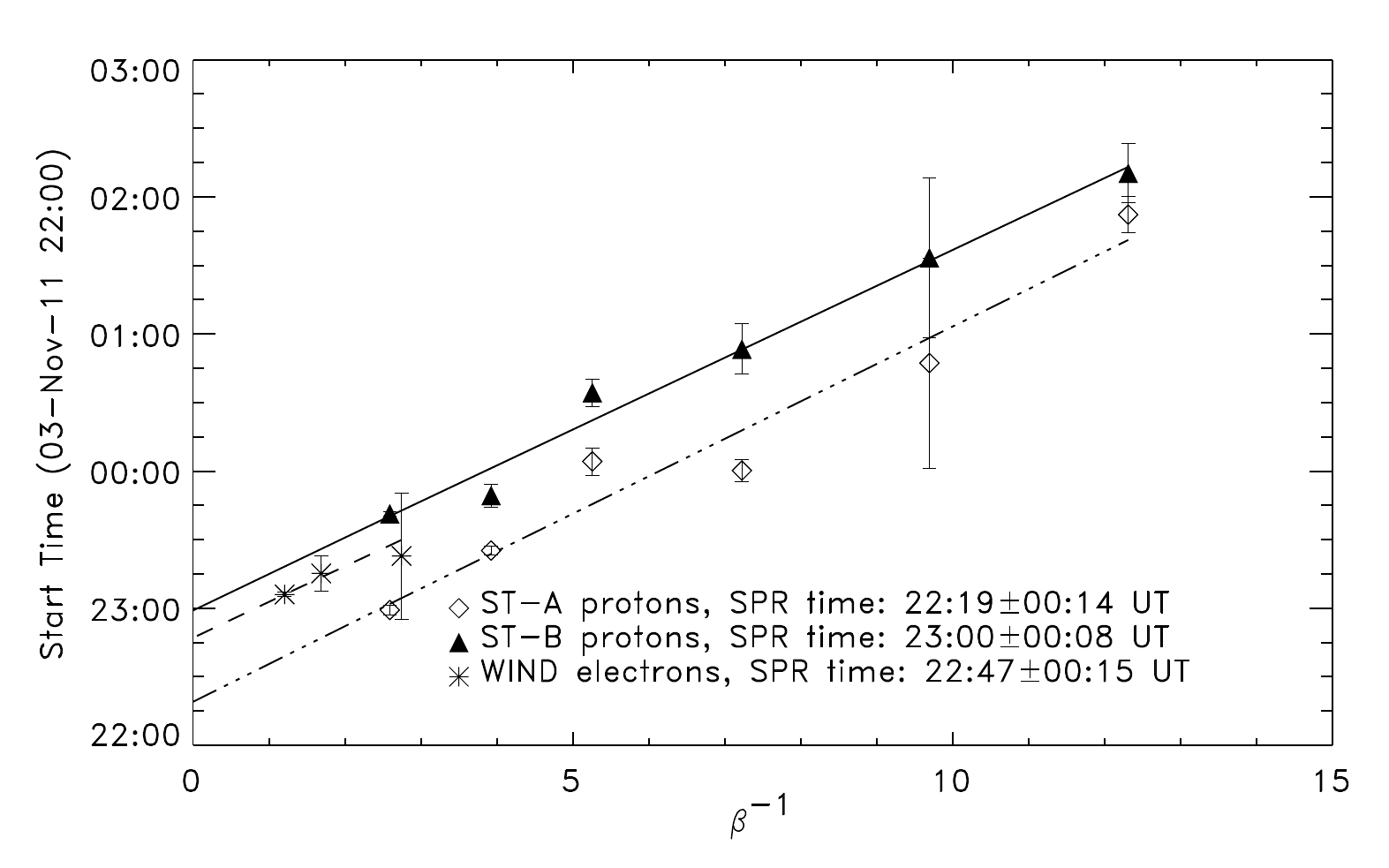}}
 \caption{Velocity dispersion of particle onset times \emph{versus} the reciprocal of the relativistic $\beta$- function ($\beta^{-1}$ = (\emph{v/c})$^{-1}$). This uses the onset of protons at STEREO-A (dot--dash line) and -B (solid line), and the onset of electrons at Wind (dashed line). The slopes of the fit lines multiplied by \emph{c} give the path lengths and the intercepts give the solar particle release (SPR) times of the particles. The error in these fits yields the uncertainty in the derived SPR times and path lengths.}\label{fig:newveldisp}
\end{figure}

Velocity-dispersion analysis was carried out on the accelerated protons detected at the STEREO spacecraft in order to determine the solar particle release (SPR) time and path length of the protons. This analysis involves plotting the onset times versus $\beta^{-1}$ = (\emph{v/c})$^{-1}$ for particles from different energy intervals. The slope of the straight fit line multiplied by \emph{c} gives the path length and the intercept gives the initial SPR time. Velocity dispersion of protons arriving at Wind was not carried out due to the fewer energy bins available; instead velocity dispersion was done on energetic electrons arriving at Wind. This is shown in Figure \ref{fig:newveldisp}. The release time of protons at STEREO-A was found to be 22:19$\pm$00:14 UT, with a path length of 1.97$\pm$0.22 AU. By comparison the release time at STEREO-B is delayed by around 50 minutes from the CME eruption until 23:00$\pm$00:08 UT, which is during the Type II radio burst seen at STEREO-B, with a path length of 1.89$\pm$0.13 AU. Velocity dispersion of electrons arriving at Wind yielded an SPR time of 22:47$\pm$00:15, with a path length of 1.86$\pm$1.44 AU.

%% Figure 
%
 \begin{figure} 
\centerline{\includegraphics[width=1.1\textwidth,clip=]{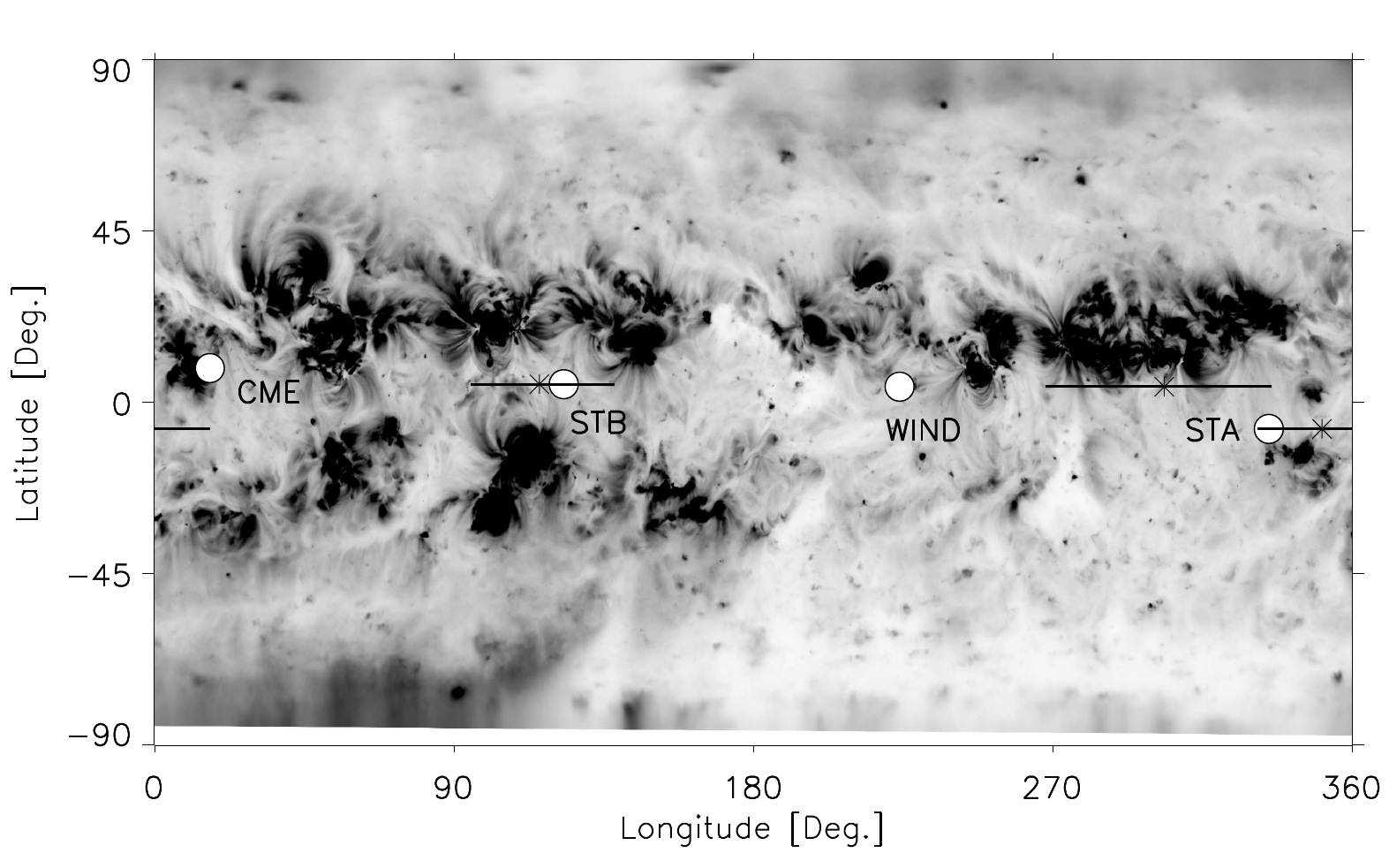}}
 \caption{A reverse-intensity Carrington map of EUV observations made in 195\,\AA\ . The magnetic footpoints of STEREO-A, STEREO-B and Wind are labelled, along with the source position of the CME (white circles). The black stars show the position of the edge of the CME, as calculated using the expansion speed from Cor-1A, at the SPR times of energetic particles in the three locations. The black lines show the uncertainty in this position, due to the uncertainty in SPR time and expansion speed.}\label{fig:carringtonplotexpspr}
\end{figure}

\section{Discussion}
By considering the expansion of the CME and the positions of the footpoints of the magnetic-field lines connecting each spacecraft to the Sun, conclusions can be drawn about the origin of the apparent delay in particle-release times and the order of the particle onsets. Consistent with the work done by \inlinecite{Rouillard2012} it is found that the release time of particles arriving at each location is dependent on the time for the CME to expand to the magnetic footpoints of each spacecraft. The locations of these magnetic footpoints were calculated using Parker spiral theory with the average solar wind speed measured \emph{in-situ} at each spacecraft during the onset of the SEP events. Those speeds were $\approx$315 kms$^{-1}$ at STEREO-A, $\approx$330 kms$^{-1}$ at STEREO-B, and $\approx$265 kms$^{-1}$ at Wind. The footpoint of STEREO-A is the closest to the erupting active region, despite this active region only being visible from STEREO-B. This can be seen from Figure \ref{fig:carringtonplotexpspr}, a reverse-intensity Carrington plot with the magnetic footpoints of STEREO-A, STEREO-B, and Wind indicated, as well as the source point of the CME (white circles as labelled). As the CME expands it will therefore reach the footpoint of STEREO-A first, so particles arriving at STEREO-A have the earliest SPR time, as is observed. The CME should reach the footpoint of STEREO-B next, as this is closer to the CME source region than the footpoint of Wind, but instead the SPR time for SEPs arriving at Wind is observed to be earlier than STEREO-B. 

In the work done by \inlinecite{Rouillard2012}, the initial expansion of the CME is approximated by the propagation of the EUV wave across the solar disk, as seen in EUV running difference images. In this study an intensity-profile method is employed, allowing for the wave to be tracked from 22:16 to 22:31 UT, which includes the particle release time of protons detected at STEREO-A. The polar plots at this height show that the CME was still expanding beyond this time however, from the initial eruption at 22:10 UT until 22:50 UT. The measured speed of the EUV wave was 221$\pm$15 kms$^{-1}$. 
At this speed the wave, and therefore the edge of the CME at this height in the corona, would not reach the magnetic footpoints of any of the spacecraft by the respective SPR times.

However, the CME did demonstrate a much faster expansion at the level of Cor-1, with the lateral expansion speed measured to be 674$\pm$38 kms$^{-1}$. The position of the edge of the CME at this height in the corona was estimated by determining the distance that the CME had expanded with this speed, in the time between the CME eruption time and the SPR time for each spacecraft (22:19, 23:00 and 22:47 UT for STEREO-A, STEREO-B, and Wind respectively). On Figure \ref{fig:carringtonplotexpspr}, the stars indicate the estimated position of the edge of the CME at each of these SPR times, using this higher expansion speed. The black lines indicate the uncertainty in this position using the uncertainty in the speed of the expansion and uncertainty in the SPR time. Figure \ref{fig:carringtonplotexpspr} shows that the edge of the CME reaches the magnetic footpoints of both of the STEREO spacecrafts at the time of its particle release within errors. This indicates that the particles detected at STEREO-A and STEREO-B were released from the Sun at the time that the edge of the CME reached the magnetic footpoints of these spacecrafts, at the height of Cor-1. The edge of the CME does not reach the footpoint connecting the Wind spacecraft to the Sun by the time of its SPR time, however.

\section{Summary and Conclusion}
The evolution of this CME was studied from several viewpoints, both on the disk and with coronagraphs. The CME erupted from the farside of the Sun, but coronagraph images show deflected streamers appearing on the opposite side of the Sun to the eruption site, indicating that the CME's influence extends to the other side of the Sun. Radio measurements at this time detected a Type II radio burst at STEREO-B, which indicates the presence of a shock in this region.

The EUV wave associated with this CME was studied using an intensity-profile technique, which only detected it for 15 minutes and determined the wave speed to be 221$\pm$15 kms$^{-1}$. The expansion of the CME across the Sun is measured using polar plots of the Sun from EUVI and Cor-1 images and demonstrates that the expansion of the CME lower down in the corona tracks the propagation of the associated EUV wave.  The speed of the CME expansion is faster higher up in the corona at the height of the polar plots from Cor-1: at 674$\pm$38 kms$^{-1}$ compared to 240$\pm$19 kms$^{-1}$ at the level of EUVI polar plots.

SEP events were detected at all three locations: STEREO-A, STEREO-B, and Wind, after the eruption of this CME, with protons and electrons arriving first at STEREO-A despite the CME being viewed on the disk from STEREO-B, but only off the limb from STEREO-A. Particle onset is then detected at Wind, followed by STEREO-B soon after. A velocity-dispersion analysis was carried out and showed that the SPR time is earliest for the particles arriving at STEREO-A (22:19$\pm$00:14 UT), then Wind (22:47$\pm$00:15 UT), and finally STEREO-B (23:00$\pm$00:08), 50 minutes after the eruption of the CME. 

The expansion of the CME is used to estimate the position of the edge of the CME on the Sun at the solar-particle release times of each spacecraft. The footpoint of the magnetic-field line connecting STEREO-A to the Sun is seen to be closer to the source location of the CME than the footpoint of STEREO-B. If the particles are released only when the CME reaches the footpoint of the spacecraft, this can explain why particles are observed at STEREO-A first.

By using the speed of the EUV wave to estimate the position of the CME, it is found that the edge of the CME does not reach the footpoints of any of the spacecraft at their respective SPR times. However, considering the faster expansion speed of the CME at the level of Cor-1 polar plots, the edge of the CME was shown to reach the footpoints of both STEREO spacecrafts at their SPR times, but not the footpoint of the Wind spacecraft located at L$_1$. This implies that for the particles arriving at the STEREO spacecrafts, the particles were released when the CME expanded higher up in the corona and reached the footpoints of these spacecrafts.
 We therefore conclude that the delay in the release time of the energetic particles at STEREO-B in relation to STEREO-A can be attributed to the time for the CME to expand out to the magnetic footpoint of STEREO-B at this higher altitude. 

This work has attempted to explain the timing of energetic particles arriving at multiple spacecrafts in relation to a CME with an associated EUV wave. \inlinecite{Rouillard2012} study a similar type of event and demonstrate that the delay in particle release in the event can be explained by the time for the CME expansion (approximated as the propagation of the EUV wave) to reach the magnetic footpoints of the spacecrafts. However, while the CME lateral expansion speed as measured in Cor-1 in this event is broadly consistent with the observed delay in particle release times, the speed of the EUV wave in the lower corona is not, and is not high enough to drive a shock in the lower corona. Also, the expansion of the CME does not reach the footpoint of Wind by 22:47 UT (the release time for particles arriving at Wind), even taking into account the uncertainty in this SPR time. This suggests that the particles observed at Wind are not associated with this CME, but originate from a different source.

More work on this event will be required to determine the cause of the energetic particles arriving at Wind. In the few hours prior to this CME, the Sun shows a lot of activity originating from a number of different regions across the Sun, that could give rise to these particles.

%%%%%%%%%%%%%%%%%%%%%%%%%%%%%%%%%%%%%%%%%%%%%%%%%%%%%%%%%%%%%%%%%%%%%%%%%%%
%% Appendix
%
% \appendix   

%%%%%%%%%%%%%%%%%%%%%%%%%%%%%%%%%%%%%%%%%%%%%%%%%%%%%%%%%%%%%%%%%%%%%%%%%%%
%% Acknowledgements
%
 \begin{acks}
We would like to thank the SOHO/LASCO, STEREO/SECCHI, IMPACT and Wind teams for providing the data used in this article. SOHO is a mission of international cooperation between ESA and NASA. We thank STFC for support via PhD studentship. The research leading to these results has received funding from the 
European Commission's Seventh Framework Programme under the grant agreement No. 284461 (eHEROES project). We would like to thank the referee, whose comments helped to significantly improve the presentation of this work.
 \end{acks}

%%% %%%%%%%%%%%%%%%%%%%%%%%%%%%%%%%%%%%%%%%%%%%%%%%%%%%%%%%%%%%
%% Bibliography
%

\end{article} 

\begin{thebibliography}{35}
% BibTex style file: spr-mp-sola.bst (nameyear), 2011-09-16
\ifx \bisbn   \undefined \def \bisbn  #1{ISBN #1}\fi
\ifx \binits  \undefined \def \binits#1{#1}\fi
\ifx \bauthor  \undefined \def \bauthor#1{#1}\fi
\ifx \batitle  \undefined \def \batitle#1{#1}\fi
\ifx \bjtitle  \undefined \def \bjtitle#1{\textit{#1}}\fi
\ifx \bvolume  \undefined \def \bvolume#1{\textbf{#1}}\fi
\ifx \byear  \undefined \def \byear#1{#1}\fi
\ifx \bissue  \undefined \def \bissue#1{#1}\fi
\ifx \bfpage  \undefined \def \bfpage#1{#1}\fi
\ifx \blpage  \undefined \def \blpage #1{#1}\fi
\ifx \burl  \undefined \def \burl#1{\textsf{#1}}\fi
\ifx \href  \undefined \def \href#1#2{\textsf{#2}}\fi
\ifx \doiurl  \undefined \def \doiurl#1{\href{http://dx.doi.org/#1}{\textsf{#1}}}\fi
\ifx \betal  \undefined \def \betal{\textit{et al.}}\fi
\ifx \binstitute  \undefined \def \binstitute#1{#1}\fi
\ifx \bctitle  \undefined \def \bctitle#1{#1}\fi
\ifx \beditor  \undefined \def \beditor#1{#1}\fi
\ifx \bpublisher  \undefined \def \bpublisher#1{#1}\fi
\ifx \bbtitle  \undefined \def \bbtitle#1{\textit{#1}}\fi
\ifx \bedition  \undefined \def \bedition#1{#1}\fi
\ifx \bseriesno  \undefined \def \bseriesno#1{\textbf{#1}}\fi
\ifx \blocation  \undefined \def \blocation#1{#1}\fi
\ifx \bsertitle  \undefined \def \bsertitle#1{\textit{#1}}\fi
\ifx \bsnm \undefined \def \bsnm#1{#1}\fi
\ifx \bsuffix \undefined \def \bsuffix#1{#1}\fi
\ifx \bparticle \undefined \def \bparticle#1{#1}\fi
\ifx \barticle \undefined \def \barticle#1{}\fi
\ifx \botherref \undefined \def \botherref#1{}\fi
\ifx \url \undefined \def \url#1{\textsf{#1}}\fi
\ifx \bchapter \undefined \def \bchapter#1{}\fi
\ifx \bbook \undefined \def \bbook#1{}\fi
\ifx \bcomment \undefined \def \bcomment#1{#1}\fi
\ifx \oauthor \undefined \def \oauthor#1{#1}\fi
\ifx \citeauthoryear \undefined \def \citeauthoryear#1{#1}\fi
\def \endbibitem {}
\ifx \bconflocation  \undefined \def \bconflocation#1{#1} \fi

\bibitem[\protect\citeauthoryear{{Bieber} \textit{et~al.}}{2005}]{Bieber2005}
\begin{barticle}
\bauthor{\bsnm{{Bieber}}, \binits{J.W.}},
\bauthor{\bsnm{{Clem}}, \binits{J.}},
\bauthor{\bsnm{{Evenson}}, \binits{P.}},
\bauthor{\bsnm{{Pyle}}, \binits{R.}},
\bauthor{\bsnm{{Ruffolo}}, \binits{D.}},
\bauthor{\bsnm{{S{\'a}iz}}, \binits{A.}}:
\byear{2005},
\batitle{{Relativistic solar neutrons and protons on 28 October 2003}}.
\bjtitle{\grl}
\bvolume{32},
\bfpage{3}.
doi:\doiurl{10.1029/2004GL021492}.
\end{barticle}
\endbibitem

\bibitem[\protect\citeauthoryear{{Bougeret}
  \textit{et~al.}}{1995}]{Bougeret1995}
\begin{barticle}
\bauthor{\bsnm{{Bougeret}}, \binits{J.-L.}},
\bauthor{\bsnm{{Kaiser}}, \binits{M.L.}},
\bauthor{\bsnm{{Kellogg}}, \binits{P.J.}},
\bauthor{\bsnm{{Manning}}, \binits{R.}},
\bauthor{\bsnm{{Goetz}}, \binits{K.}},
\bauthor{\bsnm{{Monson}}, \binits{S.J.}},
\bauthor{\bsnm{{Monge}}, \binits{N.}},
\bauthor{\bsnm{{Friel}}, \binits{L.}},
\bauthor{\bsnm{{Meetre}}, \binits{C.A.}},
\bauthor{\bsnm{{Perche}}, \binits{C.}},
\bauthor{\bsnm{{Sitruk}}, \binits{L.}},
\bauthor{\bsnm{{Hoang}}, \binits{S.}}:
\byear{1995},
\batitle{{Waves: The Radio and Plasma Wave Investigation on the Wind
  Spacecraft}}.
\bjtitle{\ssr}
\bvolume{71},
\bfpage{231}\,--\,\blpage{263}.
doi:\doiurl{10.1007/BF00751331}.
\end{barticle}
\endbibitem

\bibitem[\protect\citeauthoryear{{Bougeret}
  \textit{et~al.}}{2008}]{Bougeret2008}
\begin{barticle}
\bauthor{\bsnm{{Bougeret}}, \binits{J.L.}},
\bauthor{\bsnm{{Goetz}}, \binits{K.}},
\bauthor{\bsnm{{Kaiser}}, \binits{M.L.}},
\bauthor{\bsnm{{Bale}}, \binits{S.D.}},
\bauthor{\bsnm{{Kellogg}}, \binits{P.J.}},
\bauthor{\bsnm{{Maksimovic}}, \binits{M.}},
\bauthor{\bsnm{{Monge}}, \binits{N.}},
\bauthor{\bsnm{{Monson}}, \binits{S.J.}},
\bauthor{\bsnm{{Astier}}, \binits{P.L.}},
\bauthor{\bsnm{{Davy}}, \binits{S.}},
\bauthor{\bsnm{{Dekkali}}, \binits{M.}},
\bauthor{\bsnm{{Hinze}}, \binits{J.J.}},
\bauthor{\bsnm{{Manning}}, \binits{R.E.}},
\bauthor{\bsnm{{Aguilar-Rodriguez}}, \binits{E.}},
\bauthor{\bsnm{{Bonnin}}, \binits{X.}},
\bauthor{\bsnm{{Briand}}, \binits{C.}},
\bauthor{\bsnm{{Cairns}}, \binits{I.H.}},
\bauthor{\bsnm{{Cattell}}, \binits{C.A.}},
\bauthor{\bsnm{{Cecconi}}, \binits{B.}},
\bauthor{\bsnm{{Eastwood}}, \binits{J.}},
\bauthor{\bsnm{{Ergun}}, \binits{R.E.}},
\bauthor{\bsnm{{Fainberg}}, \binits{J.}},
\bauthor{\bsnm{{Hoang}}, \binits{S.}},
\bauthor{\bsnm{{Huttunen}}, \binits{K.E.J.}},
\bauthor{\bsnm{{Krucker}}, \binits{S.}},
\bauthor{\bsnm{{Lecacheux}}, \binits{A.}},
\bauthor{\bsnm{{MacDowall}}, \binits{R.J.}},
\bauthor{\bsnm{{Macher}}, \binits{W.}},
\bauthor{\bsnm{{Mangeney}}, \binits{A.}},
\bauthor{\bsnm{{Meetre}}, \binits{C.A.}},
\bauthor{\bsnm{{Moussas}}, \binits{X.}},
\bauthor{\bsnm{{Nguyen}}, \binits{Q.N.}},
\bauthor{\bsnm{{Oswald}}, \binits{T.H.}},
\bauthor{\bsnm{{Pulupa}}, \binits{M.}},
\bauthor{\bsnm{{Reiner}}, \binits{M.J.}},
\bauthor{\bsnm{{Robinson}}, \binits{P.A.}},
\bauthor{\bsnm{{Rucker}}, \binits{H.}},
\bauthor{\bsnm{{Salem}}, \binits{C.}},
\bauthor{\bsnm{{Santolik}}, \binits{O.}},
\bauthor{\bsnm{{Silvis}}, \binits{J.M.}},
\bauthor{\bsnm{{Ullrich}}, \binits{R.}},
\bauthor{\bsnm{{Zarka}}, \binits{P.}},
\bauthor{\bsnm{{Zouganelis}}, \binits{I.}}:
\byear{2008},
\batitle{{S/WAVES: The Radio and Plasma Wave Investigation on the STEREO
  Mission}}.
\bjtitle{\ssr}
\bvolume{136},
\bfpage{487}\,--\,\blpage{528}.
doi:\doiurl{10.1007/s11214-007-9298-8}.
\end{barticle}
\endbibitem

\bibitem[\protect\citeauthoryear{{Brueckner}
  \textit{et~al.}}{1995}]{Brueckner1995}
\begin{barticle}
\bauthor{\bsnm{{Brueckner}}, \binits{G.E.}},
\bauthor{\bsnm{{Howard}}, \binits{R.A.}},
\bauthor{\bsnm{{Koomen}}, \binits{M.J.}},
\bauthor{\bsnm{{Korendyke}}, \binits{C.M.}},
\bauthor{\bsnm{{Michels}}, \binits{D.J.}},
\bauthor{\bsnm{{Moses}}, \binits{J.D.}},
\bauthor{\bsnm{{Socker}}, \binits{D.G.}},
\bauthor{\bsnm{{Dere}}, \binits{K.P.}},
\bauthor{\bsnm{{Lamy}}, \binits{P.L.}},
\bauthor{\bsnm{{Llebaria}}, \binits{A.}},
\bauthor{\bsnm{{Bout}}, \binits{M.V.}},
\bauthor{\bsnm{{Schwenn}}, \binits{R.}},
\bauthor{\bsnm{{Simnett}}, \binits{G.M.}},
\bauthor{\bsnm{{Bedford}}, \binits{D.K.}},
\bauthor{\bsnm{{Eyles}}, \binits{C.J.}}:
\byear{1995},
\batitle{{The Large Angle Spectroscopic Coronagraph (LASCO)}}.
\bjtitle{Solar Phys.}
\bvolume{162},
\bfpage{357}\,--\,\blpage{402}.
doi:\doiurl{10.1007/BF00733434}.
\end{barticle}
\endbibitem

\bibitem[\protect\citeauthoryear{{Cane} and {Erickson}}{2003}]{Cane2003}
\begin{barticle}
\bauthor{\bsnm{{Cane}}, \binits{H.V.}},
\bauthor{\bsnm{{Erickson}}, \binits{W.C.}}:
\byear{2003},
\batitle{{Energetic particle propagation in the inner heliosphere as deduced
  from low-frequency ({\textless}100 kHz) observations of type III radio
  bursts}}.
\bjtitle{J. Geophys. Res. (Space Phys.)}
\bvolume{108},
\bfpage{1203}.
doi:\doiurl{10.1029/2002JA009488}.
\end{barticle}
\endbibitem

\bibitem[\protect\citeauthoryear{{Cliver} and {Cane}}{1996}]{Cliver1996}
\begin{barticle}
\bauthor{\bsnm{{Cliver}}, \binits{E.W.}},
\bauthor{\bsnm{{Cane}}, \binits{H.V.}}:
\byear{1996},
\batitle{{The angular extents of solar/interplanetary disturbances and
  modulation of galactic cosmic rays}}.
\bjtitle{\jgr}
\bvolume{101},
\bfpage{15533}\,--\,\blpage{15546}.
doi:\doiurl{10.1029/96JA00492}.
\end{barticle}
\endbibitem

\bibitem[\protect\citeauthoryear{{Cliver}, {Webb}, and
  {Howard}}{1999}]{Cliver1999}
\begin{barticle}
\bauthor{\bsnm{{Cliver}}, \binits{E.W.}},
\bauthor{\bsnm{{Webb}}, \binits{D.F.}},
\bauthor{\bsnm{{Howard}}, \binits{R.A.}}:
\byear{1999},
\batitle{{On the origin of solar metric type II bursts}}.
\bjtitle{Solar Phys.}
\bvolume{187},
\bfpage{89}\,--\,\blpage{114}.
\end{barticle}
\endbibitem

\bibitem[\protect\citeauthoryear{{Cliver} \textit{et~al.}}{1995}]{Cliver1995}
\begin{bchapter}
\bauthor{\bsnm{{Cliver}}, \binits{E.W.}},
\bauthor{\bsnm{{Kahler}}, \binits{S.W.}},
\bauthor{\bsnm{{Neidig}}, \binits{D.F.}},
\bauthor{\bsnm{{Cane}}, \binits{H.V.}},
\bauthor{\bsnm{{Richardson}}, \binits{I.G.}},
\bauthor{\bsnm{{Kallenrode}}, \binits{M.B.}},
\bauthor{\bsnm{{Wibberenz}}, \binits{G.}}:
\byear{1995},
\bctitle{{Extreme ''Propagation'' of Solar Energetic Particles}}.
In: \beditor{\bsnm{Iucci}, \binits{N.}},
\beditor{\bsnm{Lamanna}, \binits{E.}} (eds.)
\bbtitle{Internat. Cosmic Ray Conf.}
\bseriesno{4},
\bfpage{257}.
\end{bchapter}
\endbibitem

\bibitem[\protect\citeauthoryear{{Domingo}, {Fleck}, and
  {Poland}}{1995}]{Domingo1995}
\begin{barticle}
\bauthor{\bsnm{{Domingo}}, \binits{V.}},
\bauthor{\bsnm{{Fleck}}, \binits{B.}},
\bauthor{\bsnm{{Poland}}, \binits{A.I.}}:
\byear{1995},
\batitle{{The SOHO Mission: an Overview}}.
\bjtitle{Solar Phys.}
\bvolume{162},
\bfpage{1}\,--\,\blpage{37}.
doi:\doiurl{10.1007/BF00733425}.
\end{barticle}
\endbibitem

\bibitem[\protect\citeauthoryear{{Dresing} \textit{et~al.}}{2012}]{Dresing2012}
\begin{barticle}
\bauthor{\bsnm{{Dresing}}, \binits{N.}},
\bauthor{\bsnm{{G{\'o}mez-Herrero}}, \binits{R.}},
\bauthor{\bsnm{{Klassen}}, \binits{A.}},
\bauthor{\bsnm{{Heber}}, \binits{B.}},
\bauthor{\bsnm{{Kartavykh}}, \binits{Y.}},
\bauthor{\bsnm{{Dr{\"o}ge}}, \binits{W.}}:
\byear{2012},
\batitle{{The Large Longitudinal Spread of Solar Energetic Particles During the
  17 January 2010 Solar Event}}.
\bjtitle{Solar Phys.}
\bvolume{281},
\bfpage{281}\,--\,\blpage{300}.
doi:\doiurl{10.1007/s11207-012-0049-y}.
\end{barticle}
\endbibitem

\bibitem[\protect\citeauthoryear{{Howard} \textit{et~al.}}{2008}]{Howard2008}
\begin{barticle}
\bauthor{\bsnm{{Howard}}, \binits{R.A.}},
\bauthor{\bsnm{{Moses}}, \binits{J.D.}},
\bauthor{\bsnm{{Vourlidas}}, \binits{A.}},
\bauthor{\bsnm{{Newmark}}, \binits{J.S.}},
\bauthor{\bsnm{{Socker}}, \binits{D.G.}},
\bauthor{\bsnm{{Plunkett}}, \binits{S.P.}},
\bauthor{\bsnm{{Korendyke}}, \binits{C.M.}},
\bauthor{\bsnm{{Cook}}, \binits{J.W.}},
\bauthor{\bsnm{{Hurley}}, \binits{A.}},
\bauthor{\bsnm{{Davila}}, \binits{J.M.}},
\bauthor{\bsnm{{Thompson}}, \binits{W.T.}},
\bauthor{\bsnm{{St Cyr}}, \binits{O.C.}},
\bauthor{\bsnm{{Mentzell}}, \binits{E.}},
\bauthor{\bsnm{{Mehalick}}, \binits{K.}},
\bauthor{\bsnm{{Lemen}}, \binits{J.R.}},
\bauthor{\bsnm{{Wuelser}}, \binits{J.P.}},
\bauthor{\bsnm{{Duncan}}, \binits{D.W.}},
\bauthor{\bsnm{{Tarbell}}, \binits{T.D.}},
\bauthor{\bsnm{{Wolfson}}, \binits{C.J.}},
\bauthor{\bsnm{{Moore}}, \binits{A.}},
\bauthor{\bsnm{{Harrison}}, \binits{R.A.}},
\bauthor{\bsnm{{Waltham}}, \binits{N.R.}},
\bauthor{\bsnm{{Lang}}, \binits{J.}},
\bauthor{\bsnm{{Davis}}, \binits{C.J.}},
\bauthor{\bsnm{{Eyles}}, \binits{C.J.}},
\bauthor{\bsnm{{Mapson-Menard}}, \binits{H.}},
\bauthor{\bsnm{{Simnett}}, \binits{G.M.}},
\bauthor{\bsnm{{Halain}}, \binits{J.P.}},
\bauthor{\bsnm{{Defise}}, \binits{J.M.}},
\bauthor{\bsnm{{Mazy}}, \binits{E.}},
\bauthor{\bsnm{{Rochus}}, \binits{P.}},
\bauthor{\bsnm{{Mercier}}, \binits{R.}},
\bauthor{\bsnm{{Ravet}}, \binits{M.F.}},
\bauthor{\bsnm{{Delmotte}}, \binits{F.}},
\bauthor{\bsnm{{Auchere}}, \binits{F.}},
\bauthor{\bsnm{{Delaboudiniere}}, \binits{J.P.}},
\bauthor{\bsnm{{Bothmer}}, \binits{V.}},
\bauthor{\bsnm{{Deutsch}}, \binits{W.}},
\bauthor{\bsnm{{Wang}}, \binits{D.}},
\bauthor{\bsnm{{Rich}}, \binits{N.}},
\bauthor{\bsnm{{Cooper}}, \binits{S.}},
\bauthor{\bsnm{{Stephens}}, \binits{V.}},
\bauthor{\bsnm{{Maahs}}, \binits{G.}},
\bauthor{\bsnm{{Baugh}}, \binits{R.}},
\bauthor{\bsnm{{McMullin}}, \binits{D.}},
\bauthor{\bsnm{{Carter}}, \binits{T.}}:
\byear{2008},
\batitle{{Sun Earth Connection Coronal and Heliospheric Investigation
  (SECCHI)}}.
\bjtitle{\ssr}
\bvolume{136},
\bfpage{67}\,--\,\blpage{115}.
doi:\doiurl{10.1007/s11214-008-9341-4}.
\end{barticle}
\endbibitem

\bibitem[\protect\citeauthoryear{{Kahler}}{1994}]{Kahler1994}
\begin{barticle}
\bauthor{\bsnm{{Kahler}}, \binits{S.}}:
\byear{1994},
\batitle{{Injection profiles of solar energetic particles as functions of
  coronal mass ejection heights}}.
\bjtitle{Astrophys. J.}
\bvolume{428},
\bfpage{837}\,--\,\blpage{842}.
doi:\doiurl{10.1086/174292}.
\end{barticle}
\endbibitem

\bibitem[\protect\citeauthoryear{{Kaiser} \textit{et~al.}}{2008}]{Kaiser2008}
\begin{barticle}
\bauthor{\bsnm{{Kaiser}}, \binits{M.L.}},
\bauthor{\bsnm{{Kucera}}, \binits{T.A.}},
\bauthor{\bsnm{{Davila}}, \binits{J.M.}},
\bauthor{\bsnm{{St.~Cyr}}, \binits{O.C.}},
\bauthor{\bsnm{{Guhathakurta}}, \binits{M.}},
\bauthor{\bsnm{{Christian}}, \binits{E.}}:
\byear{2008},
\batitle{{The STEREO Mission: An Introduction}}.
\bjtitle{\ssr}
\bvolume{136},
\bfpage{5}\,--\,\blpage{16}.
doi:\doiurl{10.1007/s11214-007-9277-0}.
\end{barticle}
\endbibitem

\bibitem[\protect\citeauthoryear{{Kallenrode}
  \textit{et~al.}}{1993}]{Kallenrode1993}
\begin{barticle}
\bauthor{\bsnm{{Kallenrode}}, \binits{M.-B.}},
\bauthor{\bsnm{{Wibberenz}}, \binits{G.}},
\bauthor{\bsnm{{Kunow}}, \binits{H.}},
\bauthor{\bsnm{{M{\"u}ller-Mellin}}, \binits{R.}},
\bauthor{\bsnm{{Stolpovskii}}, \binits{V.}},
\bauthor{\bsnm{{Kontor}}, \binits{N.}}:
\byear{1993},
\batitle{{Multi-spacecraft observations of particle events and interplanetary
  shocks during November/December 1982}}.
\bjtitle{Solar Phys.}
\bvolume{147},
\bfpage{377}\,--\,\blpage{410}.
doi:\doiurl{10.1007/BF00690726}.
\end{barticle}
\endbibitem

\bibitem[\protect\citeauthoryear{{Krucker} \textit{et~al.}}{1999}]{Krucker1999}
\begin{barticle}
\bauthor{\bsnm{{Krucker}}, \binits{S.}},
\bauthor{\bsnm{{Larson}}, \binits{D.E.}},
\bauthor{\bsnm{{Lin}}, \binits{R.P.}},
\bauthor{\bsnm{{Thompson}}, \binits{B.J.}}:
\byear{1999},
\batitle{{On the Origin of Impulsive Electron Events Observed at 1 AU}}.
\bjtitle{Astrophys. J.}
\bvolume{519},
\bfpage{864}\,--\,\blpage{875}.
doi:\doiurl{10.1086/307415}.
\end{barticle}
\endbibitem

\bibitem[\protect\citeauthoryear{{Lin} \textit{et~al.}}{1995}]{Lin1995}
\begin{barticle}
\bauthor{\bsnm{{Lin}}, \binits{R.P.}},
\bauthor{\bsnm{{Anderson}}, \binits{K.A.}},
\bauthor{\bsnm{{Ashford}}, \binits{S.}},
\bauthor{\bsnm{{Carlson}}, \binits{C.}},
\bauthor{\bsnm{{Curtis}}, \binits{D.}},
\bauthor{\bsnm{{Ergun}}, \binits{R.}},
\bauthor{\bsnm{{Larson}}, \binits{D.}},
\bauthor{\bsnm{{McFadden}}, \binits{J.}},
\bauthor{\bsnm{{McCarthy}}, \binits{M.}},
\bauthor{\bsnm{{Parks}}, \binits{G.K.}},
\bauthor{\bsnm{{R{\`e}me}}, \binits{H.}},
\bauthor{\bsnm{{Bosqued}}, \binits{J.M.}},
\bauthor{\bsnm{{Coutelier}}, \binits{J.}},
\bauthor{\bsnm{{Cotin}}, \binits{F.}},
\bauthor{\bsnm{{D'Uston}}, \binits{C.}},
\bauthor{\bsnm{{Wenzel}}, \binits{K.-P.}},
\bauthor{\bsnm{{Sanderson}}, \binits{T.R.}},
\bauthor{\bsnm{{Henrion}}, \binits{J.}},
\bauthor{\bsnm{{Ronnet}}, \binits{J.C.}},
\bauthor{\bsnm{{Paschmann}}, \binits{G.}}:
\byear{1995},
\batitle{{A Three-Dimensional Plasma and Energetic Particle Investigation for
  the Wind Spacecraft}}.
\bjtitle{\ssr}
\bvolume{71},
\bfpage{125}\,--\,\blpage{153}.
doi:\doiurl{10.1007/BF00751328}.
\end{barticle}
\endbibitem

\bibitem[\protect\citeauthoryear{{Long}, {DeLuca}, and
  {Gallagher}}{2011}]{Long2011}
\begin{barticle}
\bauthor{\bsnm{{Long}}, \binits{D.M.}},
\bauthor{\bsnm{{DeLuca}}, \binits{E.E.}},
\bauthor{\bsnm{{Gallagher}}, \binits{P.T.}}:
\byear{2011},
\batitle{{The Wave Properties of Coronal Bright Fronts Observed Using
  SDO/AIA}}.
\bjtitle{Astrophys. J. Lett.}
\bvolume{741},
\bfpage{L21}.
doi:\doiurl{10.1088/2041-8205/741/1/L21}.
\end{barticle}
\endbibitem

\bibitem[\protect\citeauthoryear{{Long} \textit{et~al.}}{2011}]{Long2011a}
\begin{barticle}
\bauthor{\bsnm{{Long}}, \binits{D.M.}},
\bauthor{\bsnm{{Gallagher}}, \binits{P.T.}},
\bauthor{\bsnm{{McAteer}}, \binits{R.T.J.}},
\bauthor{\bsnm{{Bloomfield}}, \binits{D.S.}}:
\byear{2011},
\batitle{{Deceleration and dispersion of large-scale coronal bright fronts}}.
\bjtitle{Astron. Astrophys.}
\bvolume{531},
\bfpage{A42}.
doi:\doiurl{10.1051/0004-6361/201015879}.
\end{barticle}
\endbibitem

\bibitem[\protect\citeauthoryear{{Luhmann} \textit{et~al.}}{2008}]{Luhmann2008}
\begin{barticle}
\bauthor{\bsnm{{Luhmann}}, \binits{J.G.}},
\bauthor{\bsnm{{Curtis}}, \binits{D.W.}},
\bauthor{\bsnm{{Schroeder}}, \binits{P.}},
\bauthor{\bsnm{{McCauley}}, \binits{J.}},
\bauthor{\bsnm{{Lin}}, \binits{R.P.}},
\bauthor{\bsnm{{Larson}}, \binits{D.E.}},
\bauthor{\bsnm{{Bale}}, \binits{S.D.}},
\bauthor{\bsnm{{Sauvaud}}, \binits{J.-A.}},
\bauthor{\bsnm{{Aoustin}}, \binits{C.}},
\bauthor{\bsnm{{Mewaldt}}, \binits{R.A.}},
\bauthor{\bsnm{{Cummings}}, \binits{A.C.}},
\bauthor{\bsnm{{Stone}}, \binits{E.C.}},
\bauthor{\bsnm{{Davis}}, \binits{A.J.}},
\bauthor{\bsnm{{Cook}}, \binits{W.R.}},
\bauthor{\bsnm{{Kecman}}, \binits{B.}},
\bauthor{\bsnm{{Wiedenbeck}}, \binits{M.E.}},
\bauthor{\bsnm{{von Rosenvinge}}, \binits{T.}},
\bauthor{\bsnm{{Acuna}}, \binits{M.H.}},
\bauthor{\bsnm{{Reichenthal}}, \binits{L.S.}},
\bauthor{\bsnm{{Shuman}}, \binits{S.}},
\bauthor{\bsnm{{Wortman}}, \binits{K.A.}},
\bauthor{\bsnm{{Reames}}, \binits{D.V.}},
\bauthor{\bsnm{{Mueller-Mellin}}, \binits{R.}},
\bauthor{\bsnm{{Kunow}}, \binits{H.}},
\bauthor{\bsnm{{Mason}}, \binits{G.M.}},
\bauthor{\bsnm{{Walpole}}, \binits{P.}},
\bauthor{\bsnm{{Korth}}, \binits{A.}},
\bauthor{\bsnm{{Sanderson}}, \binits{T.R.}},
\bauthor{\bsnm{{Russell}}, \binits{C.T.}},
\bauthor{\bsnm{{Gosling}}, \binits{J.T.}}:
\byear{2008},
\batitle{{STEREO IMPACT Investigation Goals, Measurements, and Data Products
  Overview}}.
\bjtitle{\ssr}
\bvolume{136},
\bfpage{117}\,--\,\blpage{184}.
doi:\doiurl{10.1007/s11214-007-9170-x}.
\end{barticle}
\endbibitem

\bibitem[\protect\citeauthoryear{{Manchester}
  \textit{et~al.}}{2008}]{Manchester2008}
\begin{barticle}
\bauthor{\bsnm{{Manchester}}, \binits{W.B.} \bsuffix{IV}},
\bauthor{\bsnm{{Vourlidas}}, \binits{A.}},
\bauthor{\bsnm{{T{\'o}th}}, \binits{G.}},
\bauthor{\bsnm{{Lugaz}}, \binits{N.}},
\bauthor{\bsnm{{Roussev}}, \binits{I.I.}},
\bauthor{\bsnm{{Sokolov}}, \binits{I.V.}},
\bauthor{\bsnm{{Gombosi}}, \binits{T.I.}},
\bauthor{\bsnm{{De Zeeuw}}, \binits{D.L.}},
\bauthor{\bsnm{{Opher}}, \binits{M.}}:
\byear{2008},
\batitle{{Three-dimensional MHD Simulation of the 2003 October 28 Coronal Mass
  Ejection: Comparison with LASCO Coronagraph Observations}}.
\bjtitle{Astrophys. J.}
\bvolume{684},
\bfpage{1448}\,--\,\blpage{1460}.
doi:\doiurl{10.1086/590231}.
\end{barticle}
\endbibitem

\bibitem[\protect\citeauthoryear{{Mewaldt} \textit{et~al.}}{2008}]{Mewaldt2008}
\begin{barticle}
\bauthor{\bsnm{{Mewaldt}}, \binits{R.A.}},
\bauthor{\bsnm{{Cohen}}, \binits{C.M.S.}},
\bauthor{\bsnm{{Cook}}, \binits{W.R.}},
\bauthor{\bsnm{{Cummings}}, \binits{A.C.}},
\bauthor{\bsnm{{Davis}}, \binits{A.J.}},
\bauthor{\bsnm{{Geier}}, \binits{S.}},
\bauthor{\bsnm{{Kecman}}, \binits{B.}},
\bauthor{\bsnm{{Klemic}}, \binits{J.}},
\bauthor{\bsnm{{Labrador}}, \binits{A.W.}},
\bauthor{\bsnm{{Leske}}, \binits{R.A.}},
\bauthor{\bsnm{{Miyasaka}}, \binits{H.}},
\bauthor{\bsnm{{Nguyen}}, \binits{V.}},
\bauthor{\bsnm{{Ogliore}}, \binits{R.C.}},
\bauthor{\bsnm{{Stone}}, \binits{E.C.}},
\bauthor{\bsnm{{Radocinski}}, \binits{R.G.}},
\bauthor{\bsnm{{Wiedenbeck}}, \binits{M.E.}},
\bauthor{\bsnm{{Hawk}}, \binits{J.}},
\bauthor{\bsnm{{Shuman}}, \binits{S.}},
\bauthor{\bsnm{{von Rosenvinge}}, \binits{T.T.}},
\bauthor{\bsnm{{Wortman}}, \binits{K.}}:
\byear{2008},
\batitle{{The Low-Energy Telescope (LET) and SEP Central Electronics for the
  STEREO Mission}}.
\bjtitle{\ssr}
\bvolume{136},
\bfpage{285}\,--\,\blpage{362}.
doi:\doiurl{10.1007/s11214-007-9288-x}.
\end{barticle}
\endbibitem

\bibitem[\protect\citeauthoryear{{Muhr} \textit{et~al.}}{2011}]{Muhr2011}
\begin{barticle}
\bauthor{\bsnm{{Muhr}}, \binits{N.}},
\bauthor{\bsnm{{Veronig}}, \binits{A.M.}},
\bauthor{\bsnm{{Kienreich}}, \binits{I.W.}},
\bauthor{\bsnm{{Temmer}}, \binits{M.}},
\bauthor{\bsnm{{Vr{\v s}nak}}, \binits{B.}}:
\byear{2011},
\batitle{{Analysis of Characteristic Parameters of Large-scale Coronal Waves
  Observed by the Solar-Terrestrial Relations Observatory/Extreme Ultraviolet
  Imager}}.
\bjtitle{Astrophys. J.}
\bvolume{739},
\bfpage{89}.
doi:\doiurl{10.1088/0004-637X/739/2/89}.
\end{barticle}
\endbibitem

\bibitem[\protect\citeauthoryear{{M{\"u}ller-Mellin}
  \textit{et~al.}}{2008}]{Muller-Mellin2008}
\begin{barticle}
\bauthor{\bsnm{{M{\"u}ller-Mellin}}, \binits{R.}},
\bauthor{\bsnm{{B{\"o}ttcher}}, \binits{S.}},
\bauthor{\bsnm{{Falenski}}, \binits{J.}},
\bauthor{\bsnm{{Rode}}, \binits{E.}},
\bauthor{\bsnm{{Duvet}}, \binits{L.}},
\bauthor{\bsnm{{Sanderson}}, \binits{T.}},
\bauthor{\bsnm{{Butler}}, \binits{B.}},
\bauthor{\bsnm{{Johlander}}, \binits{B.}},
\bauthor{\bsnm{{Smit}}, \binits{H.}}:
\byear{2008},
\batitle{{The Solar Electron and Proton Telescope for the STEREO Mission}}.
\bjtitle{\ssr}
\bvolume{136},
\bfpage{363}\,--\,\blpage{389}.
doi:\doiurl{10.1007/s11214-007-9204-4}.
\end{barticle}
\endbibitem

\bibitem[\protect\citeauthoryear{{Reames}}{1993}]{Reames1993}
\begin{barticle}
\bauthor{\bsnm{{Reames}}, \binits{D.V.}}:
\byear{1993},
\batitle{{Non-thermal particles in the interplanetary medium}}.
\bjtitle{Adv. Space Res.}
\bvolume{13},
\bfpage{331}\,--\,\blpage{339}.
doi:\doiurl{10.1016/0273-1177(93)90501-2}.
\end{barticle}
\endbibitem

\bibitem[\protect\citeauthoryear{{Reames}, {Barbier}, and
  {Ng}}{1996}]{Reames1996}
\begin{barticle}
\bauthor{\bsnm{{Reames}}, \binits{D.V.}},
\bauthor{\bsnm{{Barbier}}, \binits{L.M.}},
\bauthor{\bsnm{{Ng}}, \binits{C.K.}}:
\byear{1996},
\batitle{{The Spatial Distribution of Particles Accelerated by Coronal Mass
  Ejection--driven Shocks}}.
\bjtitle{Astrophys. J.}
\bvolume{466},
\bfpage{473}.
doi:\doiurl{10.1086/177525}.
\end{barticle}
\endbibitem

\bibitem[\protect\citeauthoryear{{Rouillard}
  \textit{et~al.}}{2012}]{Rouillard2012}
\begin{barticle}
\bauthor{\bsnm{{Rouillard}}, \binits{A.P.}},
\bauthor{\bsnm{{Sheeley}}, \binits{N.R.}},
\bauthor{\bsnm{{Tylka}}, \binits{A.}},
\bauthor{\bsnm{{Vourlidas}}, \binits{A.}},
\bauthor{\bsnm{{Ng}}, \binits{C.K.}},
\bauthor{\bsnm{{Rakowski}}, \binits{C.}},
\bauthor{\bsnm{{Cohen}}, \binits{C.M.S.}},
\bauthor{\bsnm{{Mewaldt}}, \binits{R.A.}},
\bauthor{\bsnm{{Mason}}, \binits{G.M.}},
\bauthor{\bsnm{{Reames}}, \binits{D.}},
\bauthor{\bsnm{{Savani}}, \binits{N.P.}},
\bauthor{\bsnm{{StCyr}}, \binits{O.C.}},
\bauthor{\bsnm{{Szabo}}, \binits{A.}}:
\byear{2012},
\batitle{{The Longitudinal Properties of a Solar Energetic Particle Event
  Investigated Using Modern Solar Imaging}}.
\bjtitle{Astrophys. J.}
\bvolume{752},
\bfpage{44}.
doi:\doiurl{10.1088/0004-637X/752/1/44}.
\end{barticle}
\endbibitem

\bibitem[\protect\citeauthoryear{{Sheeley}, {Hakala}, and
  {Wang}}{2000}]{Sheeley2000}
\begin{barticle}
\bauthor{\bsnm{{Sheeley}}, \binits{N.R.}},
\bauthor{\bsnm{{Hakala}}, \binits{W.N.}},
\bauthor{\bsnm{{Wang}}, \binits{Y.-M.}}:
\byear{2000},
\batitle{{Detection of coronal mass ejection associated shock waves in the
  outer corona}}.
\bjtitle{\jgr}
\bvolume{105},
\bfpage{5081}\,--\,\blpage{5092}.
doi:\doiurl{10.1029/1999JA000338}.
\end{barticle}
\endbibitem

\bibitem[\protect\citeauthoryear{{Thompson}
  \textit{et~al.}}{2003}]{Thompson2003}
\begin{bchapter}
\bauthor{\bsnm{{Thompson}}, \binits{W.T.}},
\bauthor{\bsnm{{Davila}}, \binits{J.M.}},
\bauthor{\bsnm{{Fisher}}, \binits{R.R.}},
\bauthor{\bsnm{{Orwig}}, \binits{L.E.}},
\bauthor{\bsnm{{Mentzell}}, \binits{J.E.}},
\bauthor{\bsnm{{Hetherington}}, \binits{S.E.}},
\bauthor{\bsnm{{Derro}}, \binits{R.J.}},
\bauthor{\bsnm{{Federline}}, \binits{R.E.}},
\bauthor{\bsnm{{Clark}}, \binits{D.C.}},
\bauthor{\bsnm{{Chen}}, \binits{P.T.C.}},
\bauthor{\bsnm{{Tveekrem}}, \binits{J.L.}},
\bauthor{\bsnm{{Martino}}, \binits{A.J.}},
\bauthor{\bsnm{{Novello}}, \binits{J.}},
\bauthor{\bsnm{{Wesenberg}}, \binits{R.P.}},
\bauthor{\bsnm{{StCyr}}, \binits{O.C.}},
\bauthor{\bsnm{{Reginald}}, \binits{N.L.}},
\bauthor{\bsnm{{Howard}}, \binits{R.A.}},
\bauthor{\bsnm{{Mehalick}}, \binits{K.I.}},
\bauthor{\bsnm{{Hersh}}, \binits{M.J.}},
\bauthor{\bsnm{{Newman}}, \binits{M.D.}},
\bauthor{\bsnm{{Thomas}}, \binits{D.L.}},
\bauthor{\bsnm{{Card}}, \binits{G.L.}},
\bauthor{\bsnm{{Elmore}}, \binits{D.F.}}:
\byear{2003},
\bctitle{{COR1 inner coronagraph for STEREO-SECCHI}}.
In: \beditor{\bsnm{{Keil}}, \binits{S.L.}},
\beditor{\bsnm{{Avakyan}}, \binits{S.V.}} (eds.)
\bbtitle{Soc. Photo-Optical Instr. Eng. (SPIE)}
\bseriesno{CS-4853},
\bfpage{1}\,--\,\blpage{11}.
\end{bchapter}
\endbibitem

\bibitem[\protect\citeauthoryear{{Torsti}, {Riihonen}, and
  {Kocharov}}{2004}]{Torsti2004}
\begin{barticle}
\bauthor{\bsnm{{Torsti}}, \binits{J.}},
\bauthor{\bsnm{{Riihonen}}, \binits{E.}},
\bauthor{\bsnm{{Kocharov}}, \binits{L.}}:
\byear{2004},
\batitle{{The 1998 May 2-3 Magnetic Cloud: An Interplanetary ``Highway'' for
  Solar Energetic Particles Observed with SOHO/ERNE}}.
\bjtitle{Astrophys. J. Lett.}
\bvolume{600},
\bfpage{L83}\,--\,\blpage{L86}.
doi:\doiurl{10.1086/381575}.
\end{barticle}
\endbibitem

\bibitem[\protect\citeauthoryear{{Torsti} \textit{et~al.}}{1999}]{Torsti1999}
\begin{barticle}
\bauthor{\bsnm{{Torsti}}, \binits{J.}},
\bauthor{\bsnm{{Kocharov}}, \binits{L.}},
\bauthor{\bsnm{{Teittinen}}, \binits{M.}},
\bauthor{\bsnm{{Anttila}}, \binits{A.}},
\bauthor{\bsnm{{Laitinen}}, \binits{T.}},
\bauthor{\bsnm{{M{\"a}kel{\"a}}}, \binits{P.}},
\bauthor{\bsnm{{Riihonen}}, \binits{E.}},
\bauthor{\bsnm{{Vainio}}, \binits{R.}},
\bauthor{\bsnm{{Valtonen}}, \binits{E.}}:
\byear{1999},
\batitle{{Energetic (\~{}10-65 MeV) protons observed by ERNE on August 13-14,
  1996: Eruption on the solar back side as a possible source of the event}}.
\bjtitle{\jgr}
\bvolume{104},
\bfpage{9903}\,--\,\blpage{9910}.
doi:\doiurl{10.1029/1998JA900017}.
\end{barticle}
\endbibitem

\bibitem[\protect\citeauthoryear{{Tylka} \textit{et~al.}}{2003}]{Tylka2003}
\begin{bchapter}
\bauthor{\bsnm{{Tylka}}, \binits{A.J.}},
\bauthor{\bsnm{{Cohen}}, \binits{C.M.S.}},
\bauthor{\bsnm{{Dietrich}}, \binits{W.F.}},
\bauthor{\bsnm{{Krucker}}, \binits{S.}},
\bauthor{\bsnm{{McGuire}}, \binits{R.E.}},
\bauthor{\bsnm{{Mewaldt}}, \binits{R.A.}},
\bauthor{\bsnm{{Ng}}, \binits{C.K.}},
\bauthor{\bsnm{{Reames}}, \binits{D.V.}},
\bauthor{\bsnm{{Share}}, \binits{G.H.}}:
\byear{2003},
\bctitle{{Onsets and Release Times in Solar Particle Events}}.
In: \beditor{\bsnm{Kajita}, \binits{T.}},
\beditor{\bsnm{Asaoka}, \binits{Y.}},
\beditor{\bsnm{Kawachi}, \binits{A.}},
\beditor{\bsnm{Matsubara}, \binits{Y.}},
\beditor{\bsnm{Sasaki}, \binits{M.}} (eds.)
\bbtitle{Internat. Cosmic Ray Conf.}
\bseriesno{6},
\bfpage{3305}.
\end{bchapter}
\endbibitem

\bibitem[\protect\citeauthoryear{{von Rosenvinge}
  \textit{et~al.}}{1995}]{vonRosenvinge1995}
\begin{barticle}
\bauthor{\bsnm{{von Rosenvinge}}, \binits{T.T.}},
\bauthor{\bsnm{{Barbier}}, \binits{L.M.}},
\bauthor{\bsnm{{Karsch}}, \binits{J.}},
\bauthor{\bsnm{{Liberman}}, \binits{R.}},
\bauthor{\bsnm{{Madden}}, \binits{M.P.}},
\bauthor{\bsnm{{Nolan}}, \binits{T.}},
\bauthor{\bsnm{{Reames}}, \binits{D.V.}},
\bauthor{\bsnm{{Ryan}}, \binits{L.}},
\bauthor{\bsnm{{Singh}}, \binits{S.}},
\bauthor{\bsnm{{Trexel}}, \binits{H.}},
\bauthor{\bsnm{{Winkert}}, \binits{G.}},
\bauthor{\bsnm{{Mason}}, \binits{G.M.}},
\bauthor{\bsnm{{Hamilton}}, \binits{D.C.}},
\bauthor{\bsnm{{Walpole}}, \binits{P.}}:
\byear{1995},
\batitle{{The Energetic Particles: Acceleration, Composition, and Transport
  (EPACT) investigation on the WIND spacecraft}}.
\bjtitle{\ssr}
\bvolume{71},
\bfpage{155}\,--\,\blpage{206}.
doi:\doiurl{10.1007/BF00751329}.
\end{barticle}
\endbibitem

\bibitem[\protect\citeauthoryear{{von Rosenvinge}
  \textit{et~al.}}{2007}]{vonRosenvinge2007}
\begin{botherref}
\oauthor{\bsnm{{von Rosenvinge}}, \binits{T.}},
\oauthor{\bsnm{{Cummings}}, \binits{A.}},
\oauthor{\bsnm{{Cohen}}, \binits{C.}},
\oauthor{\bsnm{{Leske}}, \binits{R.}},
\oauthor{\bsnm{{Mewaldt}}, \binits{R.}},
\oauthor{\bsnm{{Stone}}, \binits{E.}},
\oauthor{\bsnm{{Wiedenbeck}}, \binits{M.}}:
2007,
{The High Energy Telescopes for the STEREO Mission}.
\textit{AGU Spring Meeting Abstracts},
A5.
\end{botherref}
\endbibitem

\bibitem[\protect\citeauthoryear{{Vourlidas}
  \textit{et~al.}}{2003}]{Vourlidas2003}
\begin{barticle}
\bauthor{\bsnm{{Vourlidas}}, \binits{A.}},
\bauthor{\bsnm{{Wu}}, \binits{S.T.}},
\bauthor{\bsnm{{Wang}}, \binits{A.H.}},
\bauthor{\bsnm{{Subramanian}}, \binits{P.}},
\bauthor{\bsnm{{Howard}}, \binits{R.A.}}:
\byear{2003},
\batitle{{Direct Detection of a Coronal Mass Ejection-Associated Shock in Large
  Angle and Spectrometric Coronagraph Experiment White-Light Images}}.
\bjtitle{Astrophys. J.}
\bvolume{598},
\bfpage{1392}\,--\,\blpage{1402}.
doi:\doiurl{10.1086/379098}.
\end{barticle}
\endbibitem

\bibitem[\protect\citeauthoryear{{Wuelser} \textit{et~al.}}{2004}]{Wuelser2004}
\begin{bchapter}
\bauthor{\bsnm{{Wuelser}}, \binits{J.-P.}},
\bauthor{\bsnm{{Lemen}}, \binits{J.R.}},
\bauthor{\bsnm{{Tarbell}}, \binits{T.D.}},
\bauthor{\bsnm{{Wolfson}}, \binits{C.J.}},
\bauthor{\bsnm{{Cannon}}, \binits{J.C.}},
\bauthor{\bsnm{{Carpenter}}, \binits{B.A.}},
\bauthor{\bsnm{{Duncan}}, \binits{D.W.}},
\bauthor{\bsnm{{Gradwohl}}, \binits{G.S.}},
\bauthor{\bsnm{{Meyer}}, \binits{S.B.}},
\bauthor{\bsnm{{Moore}}, \binits{A.S.}},
\bauthor{\bsnm{{Navarro}}, \binits{R.L.}},
\bauthor{\bsnm{{Pearson}}, \binits{J.D.}},
\bauthor{\bsnm{{Rossi}}, \binits{G.R.}},
\bauthor{\bsnm{{Springer}}, \binits{L.A.}},
\bauthor{\bsnm{{Howard}}, \binits{R.A.}},
\bauthor{\bsnm{{Moses}}, \binits{J.D.}},
\bauthor{\bsnm{{Newmark}}, \binits{J.S.}},
\bauthor{\bsnm{{Delaboudiniere}}, \binits{J.-P.}},
\bauthor{\bsnm{{Artzner}}, \binits{G.E.}},
\bauthor{\bsnm{{Auchere}}, \binits{F.}},
\bauthor{\bsnm{{Bougnet}}, \binits{M.}},
\bauthor{\bsnm{{Bouyries}}, \binits{P.}},
\bauthor{\bsnm{{Bridou}}, \binits{F.}},
\bauthor{\bsnm{{Clotaire}}, \binits{J.-Y.}},
\bauthor{\bsnm{{Colas}}, \binits{G.}},
\bauthor{\bsnm{{Delmotte}}, \binits{F.}},
\bauthor{\bsnm{{Jerome}}, \binits{A.}},
\bauthor{\bsnm{{Lamare}}, \binits{M.}},
\bauthor{\bsnm{{Mercier}}, \binits{R.}},
\bauthor{\bsnm{{Mullot}}, \binits{M.}},
\bauthor{\bsnm{{Ravet}}, \binits{M.-F.}},
\bauthor{\bsnm{{Song}}, \binits{X.}},
\bauthor{\bsnm{{Bothmer}}, \binits{V.}},
\bauthor{\bsnm{{Deutsch}}, \binits{W.}}:
\byear{2004},
\bctitle{{EUVI: the STEREO-SECCHI extreme ultraviolet imager}}.
In: \beditor{\bsnm{{Fineschi}}, \binits{S.}},
\beditor{\bsnm{{Gummin}}, \binits{M.A.}} (eds.)
\bbtitle{Soc. Photo-Optical Instr. Eng. (SPIE)}
\bseriesno{CS-5171},
\bfpage{111}\,--\,\blpage{122}.
doi:\doiurl{10.1117/12.506877}.
\end{bchapter}
\endbibitem

\end{thebibliography}
\end{document}